\documentclass[referee]{aa} 
\usepackage{natbib}         
\usepackage{graphicx}                    
\usepackage{amssymb}                    
\usepackage[usenames]{color}                         
\usepackage{txfonts}

\renewcommand{\vec}[1]{ {\mathbf #1} }

\newcommand{\strtable}{\renewcommand{\arraystretch}{1.2}}

\newcommand{\p}[1]{{\color{magenta}{#1}}}

\newcommand{\eq}[1]{Eq.~(\ref{eq:#1})} 
\newcommand{\eqs}[2]{Eqs.~(\ref{eq:#1},\ref{eq:#2})} 

\newcommand{\sect}[1]{Sect.~\ref{s:#1}} 
\newcommand{\app}[1]{Appendix~\ref{s:#1}} 
\newcommand{\fig}[1]{Fig.~\ref{f:#1}} 
\newcommand{\BE}{\begin{equation}}
\newcommand{\EE}{\end{equation}}
\newcommand{\BA}{\begin{eqnarray}}
\newcommand{\EA}{\end{eqnarray}}

\newcommand{\Nabla}{\vec{\nabla}}
\newcommand{\Times}{\vec{\times}}
\newcommand{\rmd}{{\rm d}}
\newcommand{\dS}{\rmd \vec{S}}
\newcommand{\dV}{\rmd \mathcal{V}}
\newcommand{\surf}{{\partial \mathcal{V}}}
\newcommand{\vol}{\mathcal{V}}
\newcommand{\ints}{\int_{\surf}}
\newcommand{\intv}{\int_{\vol}}
\newcommand{\avfi}[1]{\langle\,|f_i#1|\,\rangle}
\newcommand{\vA}{\vec{A}}
\newcommand{\Ax}{A_{\rm x}}
\newcommand{\Ay}{A_{\rm y}}

\newcommand{\vB}{\vec{B}}

\newcommand{\vBp}{\vB_{\rm p}}
\newcommand{\vBps}{\vB_{\rm p,s}}
\newcommand{\vBpns}{\vB_{\rm p,ns}}

\newcommand{\vBJ}{\vec{B}_{\rm J}}
\newcommand{\vBJs}{\vec{B}_{\rm J,s}}
\newcommand{\vBJns}{\vec{B}_{\rm J,ns}}
\newcommand{\vb}{\vec{b}}

\newcommand{\bx}{b_{\rm x}}
\newcommand{\by}{b_{\rm y}}

\newcommand{\vBs}{\vB_{\rm s}}
\newcommand{\vBns}{\vB_\delta}
\newcommand{\vBt}{\vB_{\rm test}}

\newcommand{\vBts}{\vB_{\rm test,s}}
\newcommand{\vBdiv}{\vB_{\rm div}}
\newcommand{\BDD}{\vB_{\rm DD}}
\newcommand{\BDDs}{\vB_{\rm DD, \rm s}}
\newcommand{\BTD}{\vB_{\rm TD}}
\newcommand{\BTDs}{\vB_{\rm TD, \rm s}}
\newcommand{\BMHD}{\vB_{\rm MHD}}
\newcommand{\BMHDs}{\vB_{\rm MHD, \rm s}}
\newcommand{\BEXsq}{\vB_{\rm Ex1}}
\newcommand{\BEXsqs}{\vB_{\rm Ex1, \rm s}}
\newcommand{\BEXfe}{\vB_{\rm Ex2}}
\newcommand{\BEXfes}{\vB_{\rm Ex2, \rm s}}
\newcommand{\BEXfep}{\vB_{\rm Ex2PP}}
\newcommand{\BEXfeps}{\vB_{\rm Ex2PP, \rm s}}

\newcommand{\curlA}{\Nabla \times \vA}

\newcommand{\divB}{\Nabla \cdot \vB}

\newcommand{\divBt}{\Nabla \cdot \vBt}
\newcommand{\Ep}{E_{\rm p}}
\newcommand{\EJ}{E_{\rm J}}
\newcommand{\Eps}{E_{\rm p,s}}
\newcommand{\EJs}{E_{\rm J,s}}
\newcommand{\Emix}{E_{\rm mix}}
\newcommand{\EdivBJ}{E_{\rm J,ns}}
\newcommand{\EdivBp}{E_{\rm p,ns}}
\newcommand{\En}{\tilde{E}}
\newcommand{\Epsn}{\En_{\rm p,s}}
\newcommand{\EJsn}{\En_{\rm J,s}}
\newcommand{\Emixn}{\En_{\rm mix}}
\newcommand{\EdivBJn}{\En_{\rm J,ns}}
\newcommand{\EdivBpn}{\En_{\rm p,ns}}

\newcommand{\Es}{E_{\rm test,s}}
\newcommand{\Ens}{E_{\rm div}}
\newcommand{\Esns}{E_{\rm s,div}}

\newcommand{\vn}{\vec{n}}
\newcommand{\vx}{\vec{x}}
\newcommand{\vy}{\vec{y}}
\newcommand{\vz}{\vec{z}}
\newcommand{\hatn}{\hat{\vn}}
\newcommand{\hatx}{\hat{\vx}}
\newcommand{\haty}{\hat{\vy}}
\newcommand{\hatz}{\hat{\vz}}
\newcommand{\Ps}{\Phi_{\surf}}

\newcommand{\eg}{\textit{e.g.}, }
\newcommand{\ie}{\textit{i.e.}, }

\begin{document} 

\title{Accuracy of magnetic energy computations}

\author{
G.~Valori\inst{1,2}  \and  P.~D\'emoulin\inst{1}  \and  E.~Pariat\inst{1} 
\and S.~Masson\inst{3} 
       }
       
\institute{
$^{1}$ LESIA, Observatoire de Paris, CNRS, Universit\'e Pierre et Marie Curie, Universit\'e Denis Diderot, 92195 Meudon, France \email{gherardo.valori@obspm.fr}\\
$^{2}$ University of Potsdam, Institute of Physics and Astronomy, 14476 Potsdam, Germany\\
$^{3}$ Space Weather Laboratory, NASA Goddard Space Flight Center, Greenbelt, MD, USA \\
}

\date{Received ***; accepted ***}

   \abstract
   {For magnetically driven events, the magnetic energy of the system is the prime energy reservoir that fuels the dynamical evolution.  
In the solar context, the free energy (\ie the energy in excess of the potential field energy)  is one of the main indicators used in space weather forecasts to predict the eruptivity of active regions.
A trustworthy estimation of the magnetic energy is therefore needed in three-dimensional (3D) models of the solar atmosphere, \eg in coronal fields reconstructions or numerical simulations.     
} 
{The expression of the energy of a system as the sum of its potential energy and its free energy (Thomson's theorem) is strictly valid when the magnetic field is exactly solenoidal. For numerical realizations on a discrete grid, this property may be only approximately fulfilled. We show that the imperfect solenoidality  induces terms in the energy that can lead to misinterpreting the amount of free energy present in a magnetic configuration.}
   {We consider a decomposition of the energy in solenoidal and nonsolenoidal parts which allows the unambiguous estimation of the nonsolenoidal contribution to the energy.  
   We apply this decomposition to six typical cases broadly used in solar physics. We quantify to what extent the Thomson theorem is not satisfied when approximately solenoidal fields are used. 
}
{
The quantified errors on energy vary from negligible to significant errors, depending on the extent of the nonsolenoidal component of the field. We identify the main source of errors and analyze the implications of adding a variable amount of divergence to various solenoidal fields. Finally, we present pathological unphysical situations where the estimated free energy would appear to be negative, as found in some previous works, and we identify the source of this error to be the presence of a finite divergence. 
}
   {We provide a method of quantifying the effect of a finite divergence in numerical fields, together with detailed diagnostics of its sources. We also compare the efficiency of two divergence-cleaning techniques.
These results are applicable to a broad range of numerical realizations of magnetic fields.}


   \maketitle

\section{Introduction}\label{s:intro}
Many astrophysical phenomena, such as stellar and solar jets, flares, and coronal mass ejections, are driven magnetically \citep[\eg][ and references therein]{Tajima02,Schrijver08}.
Magnetically dominated plasmas are systems where the long-range, magnetic interaction dominates other forces, \eg plasma pressure and gravitational forces. 
A typical example is the low-corona \citep[\eg][]{Priest03,Golub09}. 
There, the amount of energy associated with the magnetic field is much larger than other energy sources, and the dynamics of the coronal configuration is determined by the evolution of its magnetic field \citep[\eg][]{2000JGR...10523153F}.
This includes solar flares, where large currents develop in relatively small volumes \citep[\eg][]{Shibata11, Aulanier12}, and coronal mass ejections (CMEs), which are powerful expulsions of coronal material that change the local configuration of the magnetic field drastically \citep[\eg][]{2000JGR...10523153F,Amari03,Fan10}.
In the coronal plasma, the magnetic energy is therefore the prime energy reservoir that fuels the dynamical evolution of these events.

However, not all the magnetic energy is available for conversion into other forms of energy. 
Without changing the field significantly at the boundaries of the considered volume, the energy that can be converted into kinetic and thermal energies is given by the free energy, \ie by the difference between the total magnetic energy and the energy of the corresponding current-free (potential) field.
This very general result is known as Thomson's theorem, and it is based on the decomposition of the field into the sum of a current-carrying and a potential part. 
It does not depend on the presence of other forces, and is valid at any instant in time. 

The separation in the potential and free energies of Thomson's theorem is especially relevant for systems like the low-coronal field, that have different evolution time scales, as follows.
The time scale of the coronal potential field is determined by the underlying photosphere, which is an inertia-dominated plasma, unlike the corona.
This implies that the magnetic field at the photosphere has an evolution time scale that is much longer than the coronal one and that it is relatively insensitive to coronal changes.
Since the magnetic field at the photosphere largely determines the coronal field's current-free component, the latter also evolves on the long photospheric time scale. 
As a consequence of Thomson's theorem, relatively fast events, such as flares and CMEs, can only be powered by converting part of the magnetic free energy \citep[e.g.][]{Aulanier10,Karpen12}. 

In other words, the magnetic free energy is a sufficient condition for triggering active events, and  it is considered in the forecast of eruptions in the space weather context \citep[see, \eg][]{2006SSRv..123..251F,2011LRSP....8....1C}.
Therefore, in this and similar applications, an accurate estimation of the free energy is paramount for understanding the observed magnetic field dynamics and the maximum energy that can be released in a flare or in a CME \citep[][ and references therein]{Emslie12,Aulanier13}.

On the other hand, the free energy only provides an upper limit to the energy available for coronal dynamics. 
For instance, in the case of a flare/eruption, the post-event magnetic field configuration does not need to be potential \citep[see, \eg][]{1985ApJS...59..433B,1986RvMP...58..741T,2001JGR...10625141L}.
Indeed, flare (reconnected) loops are frequently observed to be sheared after a flare/eruption \citep[see \eg][and references therein]{Asai2003,Lin2010,Savage2012}, a feature that is also reproduced in numerical simulations \citep{Aulanier12}.
This is an indication that post-event configurations have finite free energy, and the actual energy removed by the event is given by the difference between the free energy of the pre- and post-event configurations. 
An assessment of the true energy budget related to a flare/eruptive event requires an accurate and reliable estimation of the magnetic energy.

Another motivation of this study is to address the occurrence of unphysical magnetic configurations.
This is the case in some nonlinear force-free field (NLFFF) extrapolations when nonpreprocessed, observed  vector magnetograms are used as boundary conditions. 
The most obvious evidence of the nonphysical nature of some solutions is when the energy of the extrapolated field is lower than the potential field energy.
This happens, for instance, in some of the solutions given in Table~3 of \cite{2008SoPh..247..269M} and Table~1 of \cite{2008ApJ...675.1637S} for three of the considered extrapolation methods, including one used in the present manuscript \citep{2010A&A...519A..44V}.
More generally, for all methods, the estimated coronal energy depends on the manipulations performed on the observed data prior to their use in the actual extrapolation. 
(This step is called preprocessing, \citeauthor{2006SoPh..233..215W} \citeyear{2006SoPh..233..215W}, \citeauthor{2007A&A...476..349F} \citeyear{2007A&A...476..349F}.) 
A significant part of the energy difference can eventually result from the details of the undergone preprocessing.

As a result, the understanding of basic physical processes in the solar atmosphere requires an accurate estimations of the magnetic free energy.
On the other hand, coronal models like NLFFF extrapolations, have shown that such accurate estimations are not easily obtained.
In such cases, Thomson's theorem can be exploited to address the accuracy of (free) energy  estimations. 
The fundamental assumption in Thomson's theorem is that the magnetic field is solenoidal. 
Such a property is only approximately fulfilled in numerical simulations and, more generally, in magnetic fields that are discretized on a mesh.
A quantitative estimation of the effects caused by nonvanishing field divergence is complicated by its nonlocal nature. 

The main aim of this article is to quantify the effect of the presence of a nonsolenoidal component on the energy of a discretized magnetic field. 
This is studied using six different test magnetic fields that are a sample of the typical and characteristic examples used in the context of coronal solar physics.
In the first part of the article, the energy of each test field is decomposed and interpreted using an extension of Thomson's theorem that can be applied to nonsolenoidal fields. 
In the second part we study how the energy changes, starting from a solenoidal version of each test field and adding a parametric divergent component. 
The method and results of this study are of interest when working with any discretization of magnetic fields, \eg  for 3D coronal magnetic field  extrapolations, as well as for magneto-hydrodynamic (MHD) simulations. 

In \sect{solfields} the Thomson theorem for the energy of a magnetic field is summarized. 
The extension to nonsolenoidal, discretized fields is presented in \sect{nonsolfields}. 
Section~\ref{s:testfields} introduces the six discretized fields together with their corresponding solenoidal versions that are used as test cases for applying  our analysis, whose results are given in \sect{test_Thompson}. 
Possible source of errors in our analysis are sort out in \sect{source_errors}. 
Then, in \sect{Parametric} we present the parametric study of the energy dependence on the amount of divergence added to solenoidal magnetic fields.
An analysis specific to numerical fields obtained by NLFFF extrapolations of observed vector magnetograms is presented in \sect{divestrap}, and  conclusions are finally given  in \sect{conclusions}.
\section{Magnetic energy of solenoidal fields} \label{s:solfields}
We first consider the decomposition of the magnetic energy for perfectly solenoidal fields.
By decomposing the field $\vB$ as the sum of a potential, $\vBp=\Nabla \phi$, and a current carrying contribution, $\vBJ$, 
\BE 
\vB=\vBp+\vBJ \, , \nonumber
\EE
the total magnetic energy $E$ in CGS-Gaussian units in a volume $\vol$ is given by
\BA  
  E &\equiv& \frac{1}{8\pi}\intv dV\ B^2  \nonumber \\
    & =    & \Ep+\EJ +\frac{1}{4\pi} \ints (\phi \vBJ) \cdot \dS  - \frac{1}{4\pi} \intv \phi (\Nabla \cdot \vBJ) \ \dV \,,
                          \label{eq:kel_all}
  \EA
where 
\BE
    \Ep\equiv \frac{1}{8\pi}\intv B_{\rm p}^2 \ \dV \,, \qquad 
   \EJ \equiv \frac{1}{8\pi}\intv  B_{J}^2 \ \dV \,, 
 \nonumber 
\EE
$\surf$ represents the boundary of $\vol$, $\dS=\hatn~\rmd S$, and  $\hatn$ is the external normal to the bounding surface.

Two  conditions are classically considered:
\renewcommand{\theenumi}{\alph{enumi}}
\renewcommand{\labelenumi}{[\theenumi]} 
\begin{enumerate}
   \item \label{conditiona}  $\hatn \cdot (\vB-\vBp)|_{\surf}=0$, \ie the potential field $\vBp$ is computed from the same distribution of normal field of $\vB$ on the boundary of $\vol$. This condition implies that   $\hatn \cdot \vBJ|_{\surf}=0$ and  the surface integral vanishes in \eq{kel_all};
   \item \label{conditionb} $\Nabla \cdot \vBJ=0$, in which case also the rightmost volume integral in \eq{kel_all} vanishes.
\end{enumerate}
If these two conditions hold, then
  \BE 
  E=\Ep+\EJ,
  \label{eq:kelvin_exact}
  \EE
and the energy of a magnetic field is bounded from below by the energy of the corresponding potential field that has the same distribution of the normal component on the boundary of the considered volume.
When applied to discretized fields, the above result holds under the implicit assumption that fields are numerically well resolved, yielding, in particular, continuous derivatives. 

The mathematical equivalent of \eq{kelvin_exact} is known as Thomson's (or Dirichlet's) theorem, see \eg \cite{lawrence1998}.

To satisfy the above requirement [\ref{conditiona}], the scalar potential $\phi (x,y,z)$ is computed as the solution of the Laplace equation
  \BE 
  \left \{ \begin{array}{l} \Delta \phi=0 \\
    ({\partial \phi}/{\partial \hat{n}}) |_{\surf} 
   = (\hatn \cdot \vB)|_{\surf}
  \end{array} \right. .
  \label{eq:laplace}
  \EE
In practical applications, \eq{laplace} can be solved numerically using standard methods. 
In the applications presented in this paper, the Poisson solver included in the Intel\textsuperscript{\textregistered} Mathematical Kernel Library was used.
\section{Magnetic energy of nonsolenoidal fields} \label{s:nonsolfields}
In this section we provide expressions for evaluating errors in the energy that stem from an imperfect fulfillment of the solenoidal property, as is the case for discretized magnetic fields. 
In deriving \eq{kel_all} the divergence theorem, \ie 
  \BE 
  \intv \Nabla \cdot \vec{u} \  \dV =\ints  \vec{u}  \cdot \dS \,,
  \label{eq:divTheorem}
  \EE
is used, which may not be fulfilled by the techniques employed in constructing the numerical representations of magnetic fields or in their analysis.
Moreover, if the numerically computed potential field $\vBp$ and  current-carrying field $\vBJ$  have a finite divergence, additional contributions can appear in the corresponding energy terms, $\Ep$ and $\EJ$.

We, therefore, seek a formulation of \eq{kel_all} for applications to numerical, nonsolenoidal fields that includes all possible sources of errors separately, that satisfies the requirement [\ref{conditiona}], and that includes only volume integrals (thus avoiding using the divergence theorem). 
To obtain that, we first introduce the method of computing the potential and current-carrying parts.
\subsection{Helmholtz decomposition of the potential part of the field} \label{s:potential}
The accuracy in the numerical solution of \eq{laplace} is limited, which may result in a finite divergence of the potential field. 
To quantify its effect, we can write
  \BE 
\vBp= \vB_{\rm p,s} +\Nabla \zeta \,, \qquad \mathrm{where} \qquad
  \left \{ \begin{array}{l} \Delta \zeta=\Nabla \cdot \vBp \\
    ({\partial \zeta}/{\partial \hat{n}}) |_{\surf}
   = 0 
  \end{array} \right. ,
  \label{eq:splitp} \\
  \EE
which separates in $\vBp$ the solenoidal part, $\vB_{\rm p,s}\equiv\vBp-\vB_{\rm p,ns}$ from  the nonsolenoidal one, $\vB_{\rm p,ns}\equiv\Nabla \zeta$.
This is equivalent to adopting the Helmholtz decomposition for the vector $\vBp$, together with the choice that all the nonsolenoidal component of $\vBp$ is contained in $\Nabla \zeta$.
Finally, the boundary condition for $\zeta(x,y,z)$ in \eq{splitp} is chosen such that $\vB_{\rm p,s}$ satisfies the same boundary condition as $\vBp$; \ie they both fulfill requirement [\ref{conditiona}]. 

In practical applications, we first solve \eq{laplace} numerically to determine $\phi$, then we compute $\vBp=\Nabla \phi$, and finally we overwrite the values of the normal components of $\vBp$ on each boundary according to \eq{laplace}.
Since the latter operation enforces the requirement [\ref{conditiona}], then any residual inaccuracy in the solution of \eq{laplace}, close to the boundary, implies a jump in the field, \ie a finite divergence that adds to the divergence of the potential field discussed above.  
Second, we solve \eq{splitp} to compute the residual nonsolenoidal component in $\vBp$.

\subsection{Helmholtz decomposition of the current-carrying part of the field} \label{s:current}
Using the Helmholtz decomposition on $\vBJ$ we define a solenoidal component, $\vBJs$, and a nonsolenoidal one,  $\vB_{\rm J,ns}$, such that
\BE 
\vBJ \equiv \vBJs +\nabla \psi  \,, \qquad \mathrm{where} \qquad   \left \{ \begin{array}{l} \Delta \psi=\Nabla \cdot \vBJ \\
    ({\partial \psi}/{\partial \hat{n}}) |_{\surf} 
   = 0
  \end{array} \right. ,
  \label{eq:poisson}
\EE
the nonsolenoidal part of $\vBJ$ being: $\vB_{\rm J,ns}\equiv \Nabla \psi$.
The boundary condition for $\psi$ in \eq{poisson} is chosen to have the same boundary condition for $\vB_{\rm J,s}$ and $\vBJ$, \ie to fulfill the requirement [\ref{conditiona}].
As for the potential field, the required values of  $\vBJs$ at the boundaries (\ie zero in this case) are overwritten onto the solution of \eq{poisson} which is obtained numerically, so that any error in matching these values by $\psi(x,y,z)$ reduces to a finite jump close to the boundaries. 

Finally, we notice that this method is often used to remove the divergence of a vector field \citep[][ sometimes referred to as ``projection method'']{1980JCoPh..35..426B}, and it has the property of conserving the current, \ie $\Nabla \times \vBJ=\Nabla \times \vBJs$.

\subsection{Gauge-invariant decomposition of the magnetic energy}\label{s:decomposition}
We now summarize the procedure for the decomposition of the magnetic field.
For a given numerical magnetic field $\vB$, we solve \eq{laplace} numerically and compute the corresponding potential component $\vBp$ and current-carrying component $\vBJ=\vB-\vBp$.
Next, we compute the solenoidal component $\vBps=\vBp-\Nabla\zeta$ and the nonsolenoidal component $\vBpns=\Nabla\zeta$ of the potential field by solving \eq{splitp} numerically.
Similarly, the numerical solution of \eq{poisson} provides the solenoidal component $\vBJs=\vBJ-\Nabla\psi$ and the nonsolenoidal component $\vBJns=\Nabla\psi$ of the current-carrying part of $\vB$.
The values of the different components at the boundary are such that the condition [\ref{conditiona}] is satisfied (\sect{solfields}). 
Finally, by substituting the field decomposition in $E=\intv{B^2}\dV /8\pi$ and grouping it again as in \eq{kel_all}, we obtain
  \BE
   E = \Eps +\EJs +\EdivBp +\EdivBJ +\Emix,
  \label{eq:kelvin}
  \EE
with
  \BA
  \Eps   &=&\frac{1}{8\pi} \intv B_{\rm p,s}^2 \dV \, ,\qquad 
  \EdivBp = \frac{1}{8\pi} \intv |\Nabla\zeta|^2 \dV \, \nonumber  \\
  \EJs   &=&\frac{1}{8\pi} \intv B_{\rm  J,s}^2 \dV \, ,\qquad 
  \EdivBJ = \frac{1}{8\pi} \intv |\Nabla\psi|^2 \dV \, \nonumber   \\
  \Emix  &=&\frac{1}{4\pi} \left ( 
           \intv \vB_{\rm p,s}\cdot \Nabla\zeta   \ \dV+
           \intv \vB_{\rm J,s}\cdot \Nabla\psi    \ \dV+ \right . \nonumber \\
&&   \ \ \ \ \ \ \       \intv \vB_{\rm p,s}\cdot \Nabla\psi    \ \dV+
           \intv \vB_{\rm J,s}\cdot \Nabla\zeta   \ \dV+ \nonumber \\
&&   \ \ \ \ \ \ \ \left . \intv \Nabla\zeta  \cdot \Nabla\psi   \ \dV+
           \intv \vB_{\rm p,s}\cdot \vB_{\rm J,s} \ \dV \right ) \, . 
   \label{eq:emixprime}
  \EA
All terms in \eq{kelvin} are positively defined, except for $\Emix$.   
For a perfectly solenoidal field, it is $\Eps=\Ep$, $\EJs=\EJ$, $ \EdivBp=\EdivBJ=\Emix=0$, and \eq{kelvin} reduces to \eq{kelvin_exact}.

Finally, \eq{kelvin} is normalized such that
  \BE
   1 = \Epsn +\EJsn +\EdivBpn +\EdivBJn +\Emixn \,,
  \label{eq:kelvinn}
  \EE
where the tilde indicates that the corresponding definition in \eq{emixprime} is divided by $E$. 

Using the divergence theorem, \eq{divTheorem}, and the condition [\ref{conditiona}], several terms in the above expressions could be simplified. 
However, since practical test fields may be obtained with methods that do not insure that the divergence theorem holds numerically, we have kept all the terms in \eq{emixprime}. 
Indeed, the simplification obtained by using the divergence theorem results in mixing other numerical issues with the issue of the finite divergence, producing cumbersome results, up to the point where \eq{kelvin} is not satisfied numerically.
Moreover, the direct appearance in the integrals of the scalar potentials, rather then their gradients, introduces an undesired gauge-dependence. 

\subsection{Sources of the violation of the Thomson theorem}\label{s:errthompson}
We summarize which are the source of errors that we consider in \eq{kelvin}.
First, the energy is affected by the finite divergence of the current-carrying part of the magnetic field, which enters the $\EdivBJ$ and $\Emix$ terms.
Additionally, the potential field may have a finite divergence, owing to the limited numerical accuracy of the solution of \eq{laplace}, both in the volume and close to its boundary. 
These effects are contained in the $\EdivBp$ and $\Emix$ terms.
 
As long as these are the only source of errors, then the sum of the terms on the righthand side of \eq{kelvin} must be equal to the total energy $E$ computed using $\vB$ directly, and \eq{kelvin} must hold numerically even for nondivergence-free fields. 
Equivalently but using  normalized quantities, the sum on the righthand side of \eq{kelvinn} must be equal to one.
We show in \sect{test_Thompson} that the total energy is indeed retrieved by the decomposition we adopted, allowing us to identify the source and extent of the eventual violation of Thomson's theorem, \eq{kelvin_exact}.

\subsection{Accuracy of the decomposition of the energy equation}\label{s:accuracy}
A further step is the assessment of the accuracy of the decomposition, \eq{kelvin}. 
First, we address how effective the decomposition in the solenoidal and nonsolenoidal parts is in concrete numerical applications in \sect{divergence}. 

Second, the continuity condition, implicit in the derivation of \eq{kelvin}, implies that numerical derivatives can be computed precisely enough in the employed discretization. 
This may not be the case in some numerical applications, \eg when observed values are used as boundary conditions for computing magnetic fields. 
The continuity of the fields in relation to small scales is discussed in \sect{divestrap}.

Finally, our decomposition employs the numerical solution of Laplace and Poisson equations.
We briefly recall the conditions for uniqueness of the general Poisson equation
  \BE 
  \left \{ \begin{array}{l} \Delta u=f \\
    ({\partial u}/{\partial \hat{n}}) |_{\surf} = g 
  \end{array} \right. \,,
  \label{eq:genpoiss}
  \EE
where $f(x,y,z)$ is a source term in $\vol$, and $g$ is the boundary value on $\surf$.
The use of Neumann boundary conditions implies that the solution $u(x,y,z)$ is only unique up to an additive constant.
For  Eqs.~(\ref{eq:laplace}), (\ref{eq:splitp}) and (\ref{eq:poisson}), the freedom in the additive constant is equivalent to a gauge freedom for the scalar potentials $\phi$, $\zeta$, and $\psi$, respectively.  
This gauge dependence is, however, irrelevant for \eq{kelvin}, since the energy decomposition is intentionally derived in a way such that the scalar potentials only appear in conjunction with the gradient operator.

Integrating \eq{genpoiss} in $\vol$ and using the divergence theorem, \eq{divTheorem}, we find that source and boundary values must satisfy
\BE
\ints g  \, = \intv f  \, , \label{eq:gen_poisson}
\EE
which is a necessary condition for the uniqueness of the solution $u$.
This implies that, for \eq{laplace} where $f=0$ and $g=\hat{\vn}\cdot \vB |_{\surf}$, the flux of $\vB$ through $\surf$ must vanish. 
For \eq{splitp}, where $f=\Nabla \cdot \vBp$ and $g=0$, it implies that the volume integral of $\Nabla \cdot \vBp$ must vanish. 
Similarly, for \eq{poisson}, where $f=\Nabla \cdot \vBJ$ and $g=0$, the volume integral of $\Nabla \cdot \vBJ$ must vanish. 
When such conditions cannot be insured, the uniqueness of the solution is not guaranteed.
The effect of the violation of \eq{gen_poisson} is studied in \sect{imbalance}.

\section{Test fields}\label{s:testfields}
 \begin{figure*}
 \centerline{
   \includegraphics[width=0.33\textwidth]{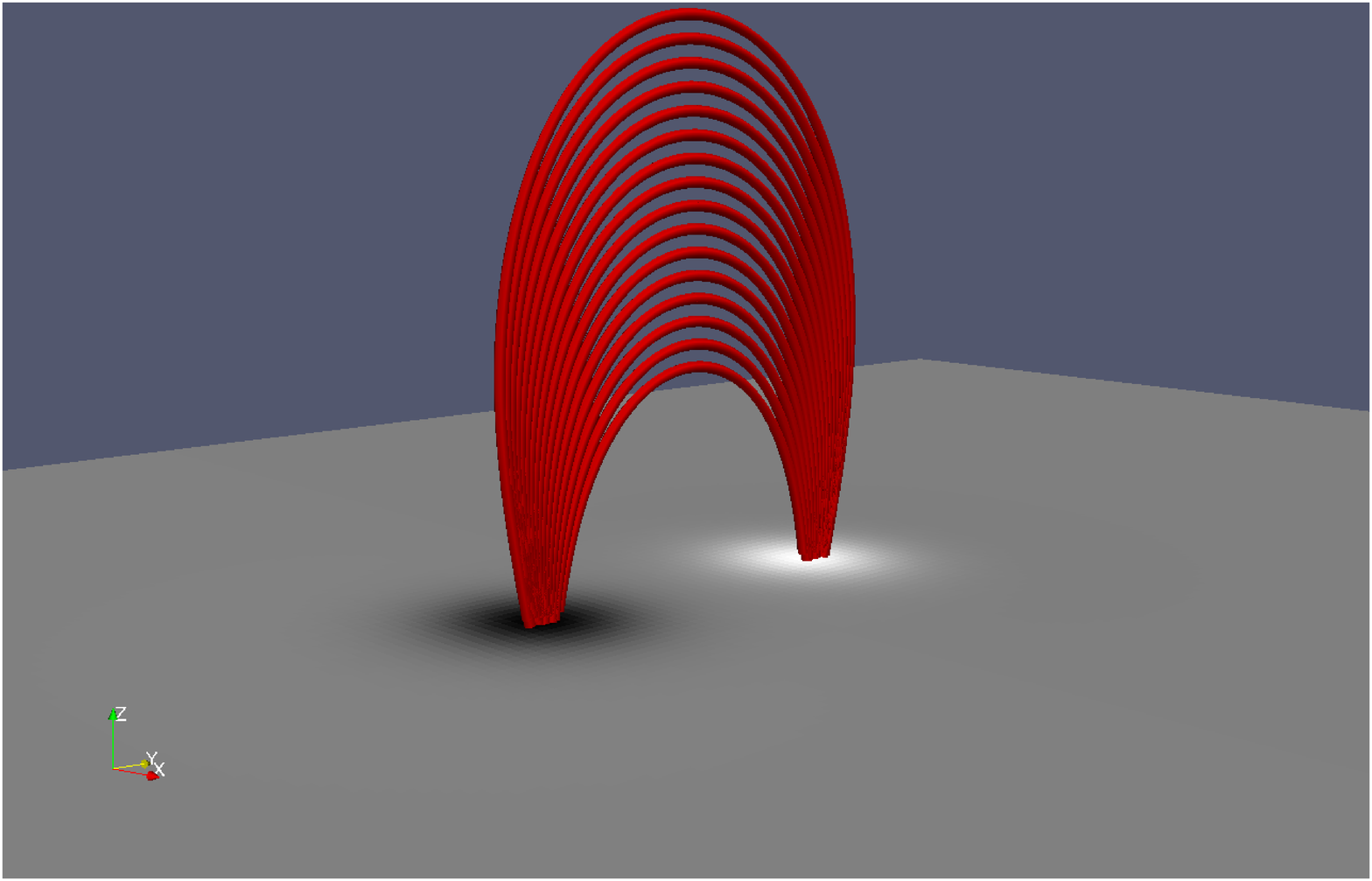}
   \includegraphics[width=0.33\textwidth]{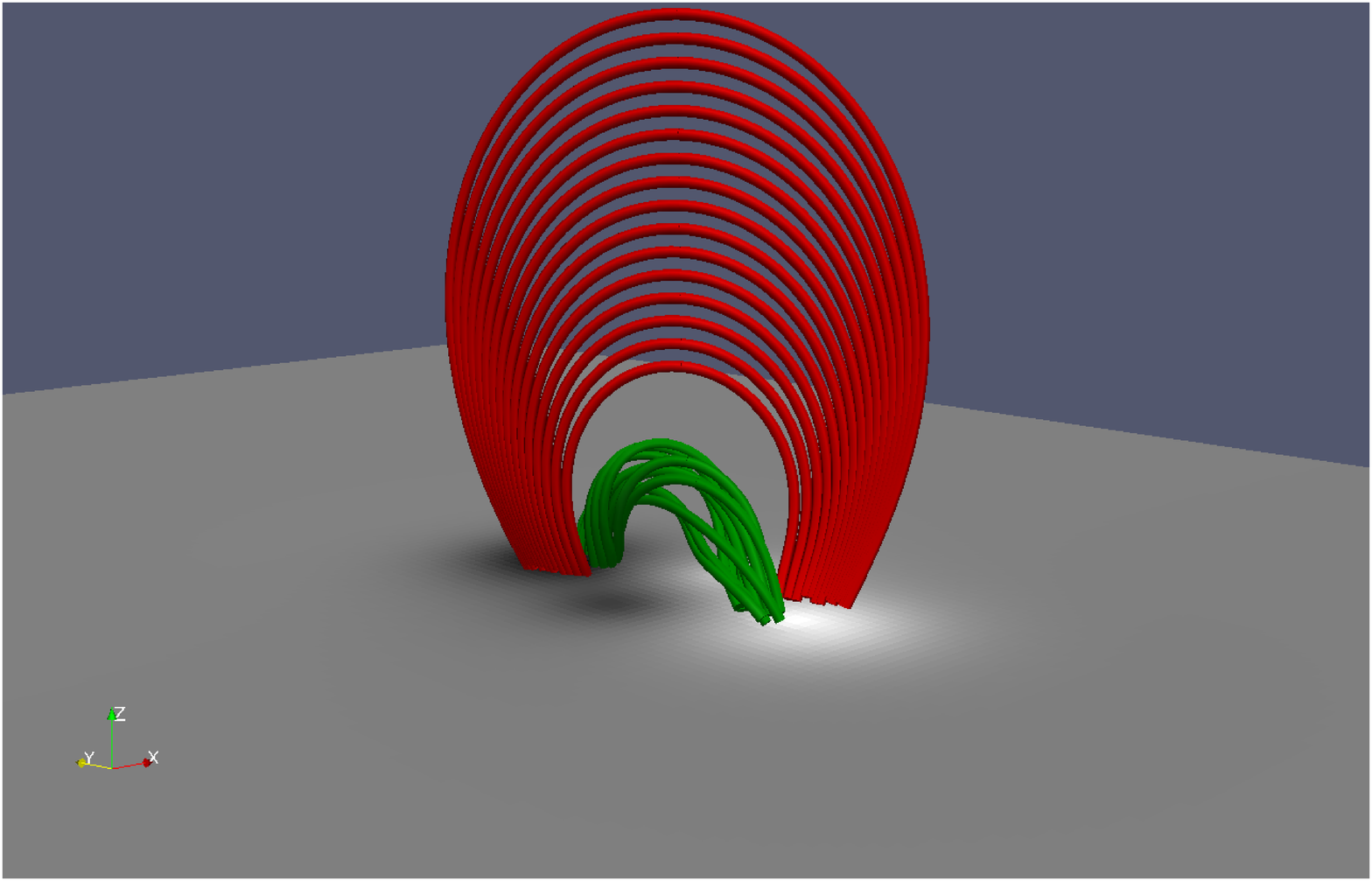}
   \includegraphics[width=0.33\textwidth]{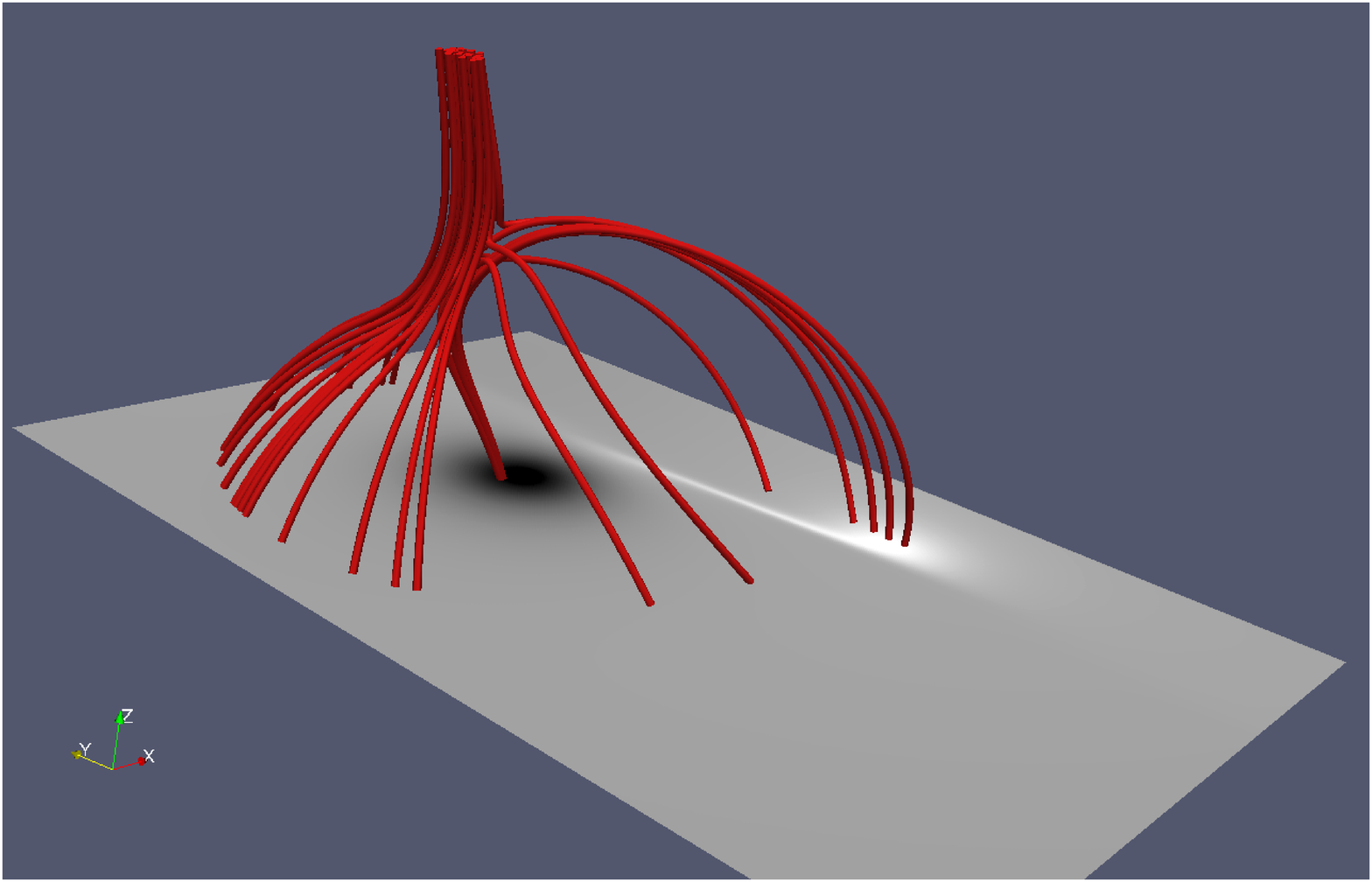}
                }
 \vspace{1mm}
  \centerline{
   \includegraphics[width=0.33\textwidth]{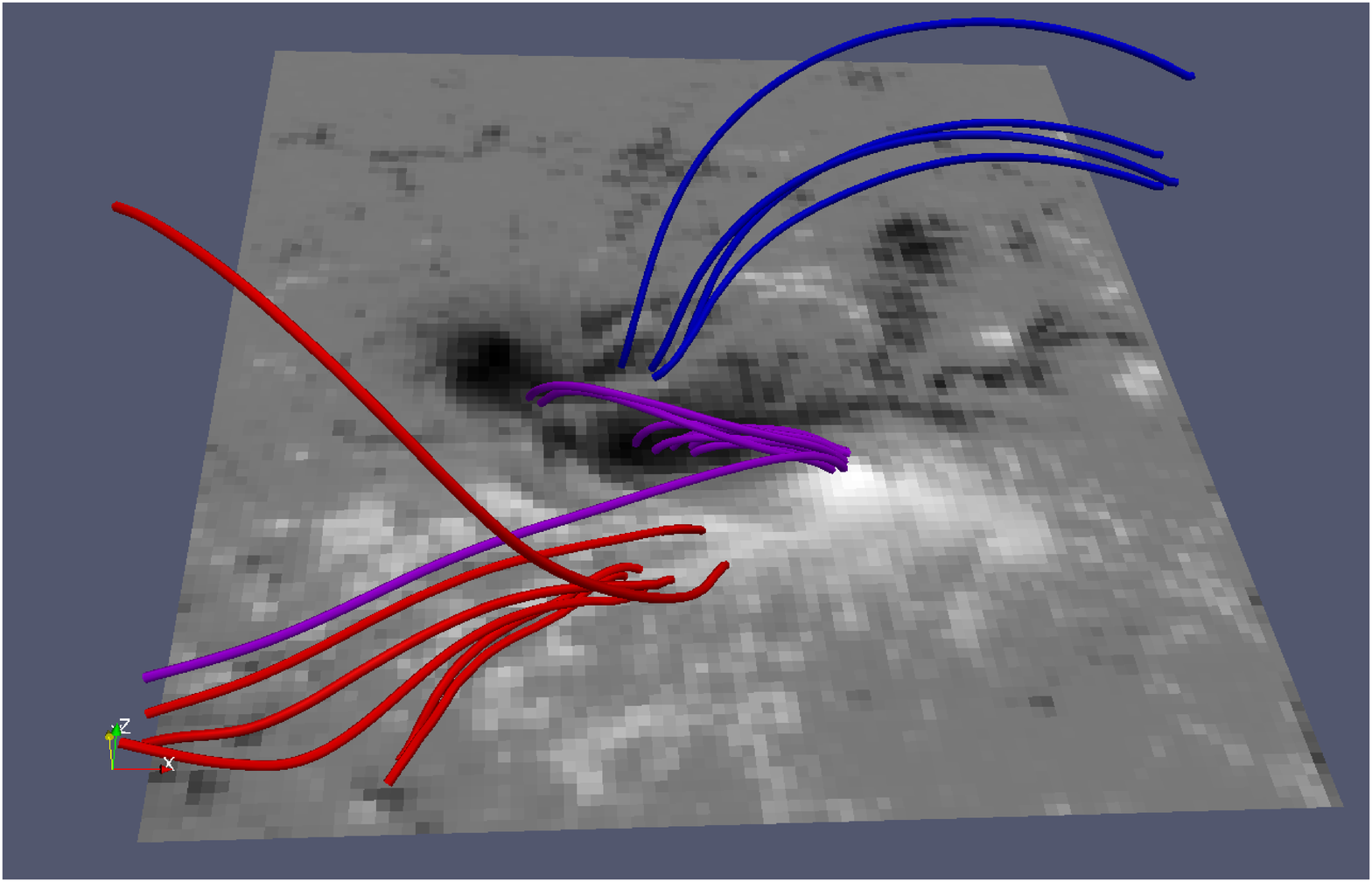}
   \includegraphics[width=0.33\textwidth]{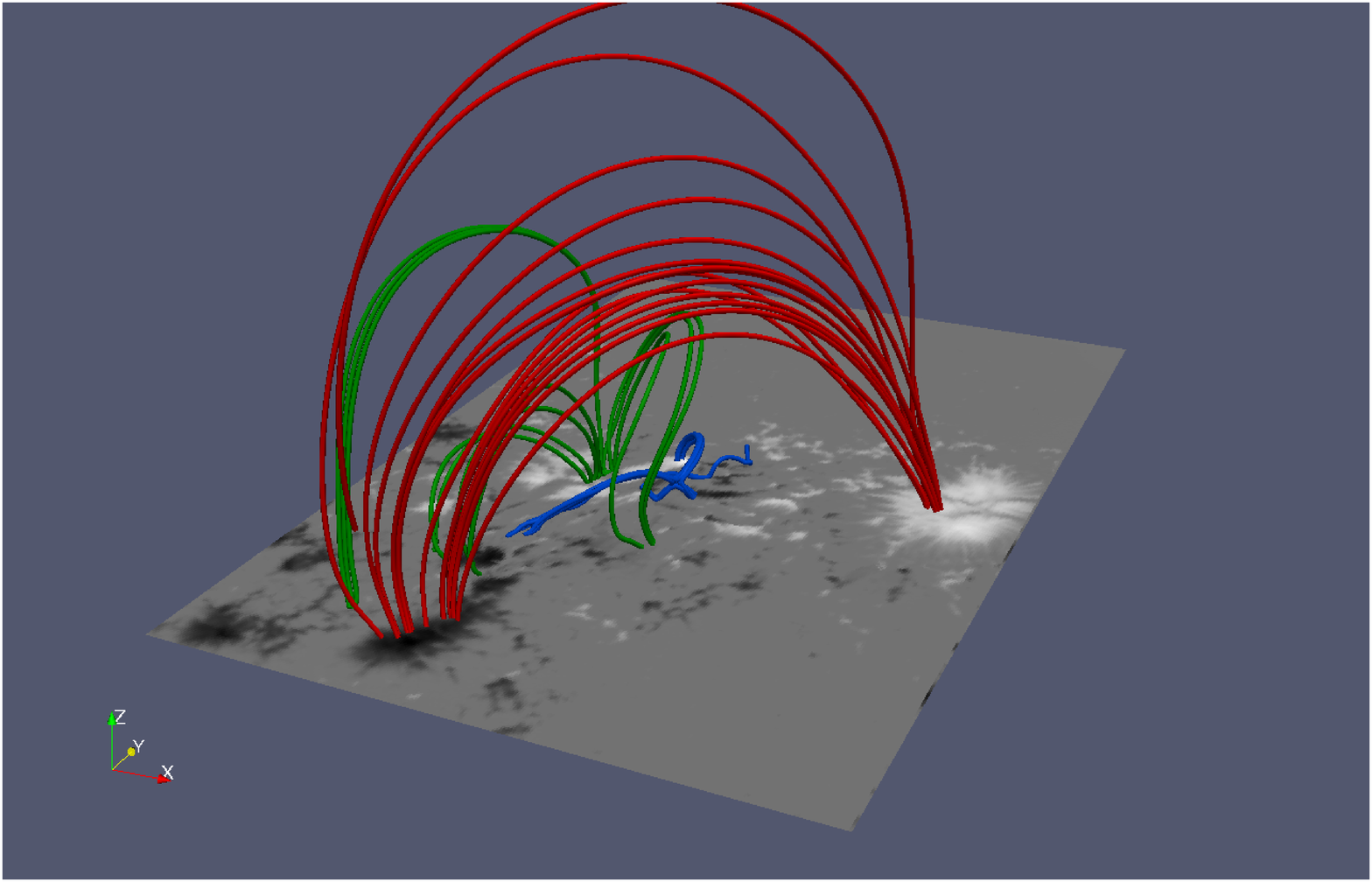}
   \includegraphics[width=0.33\textwidth]{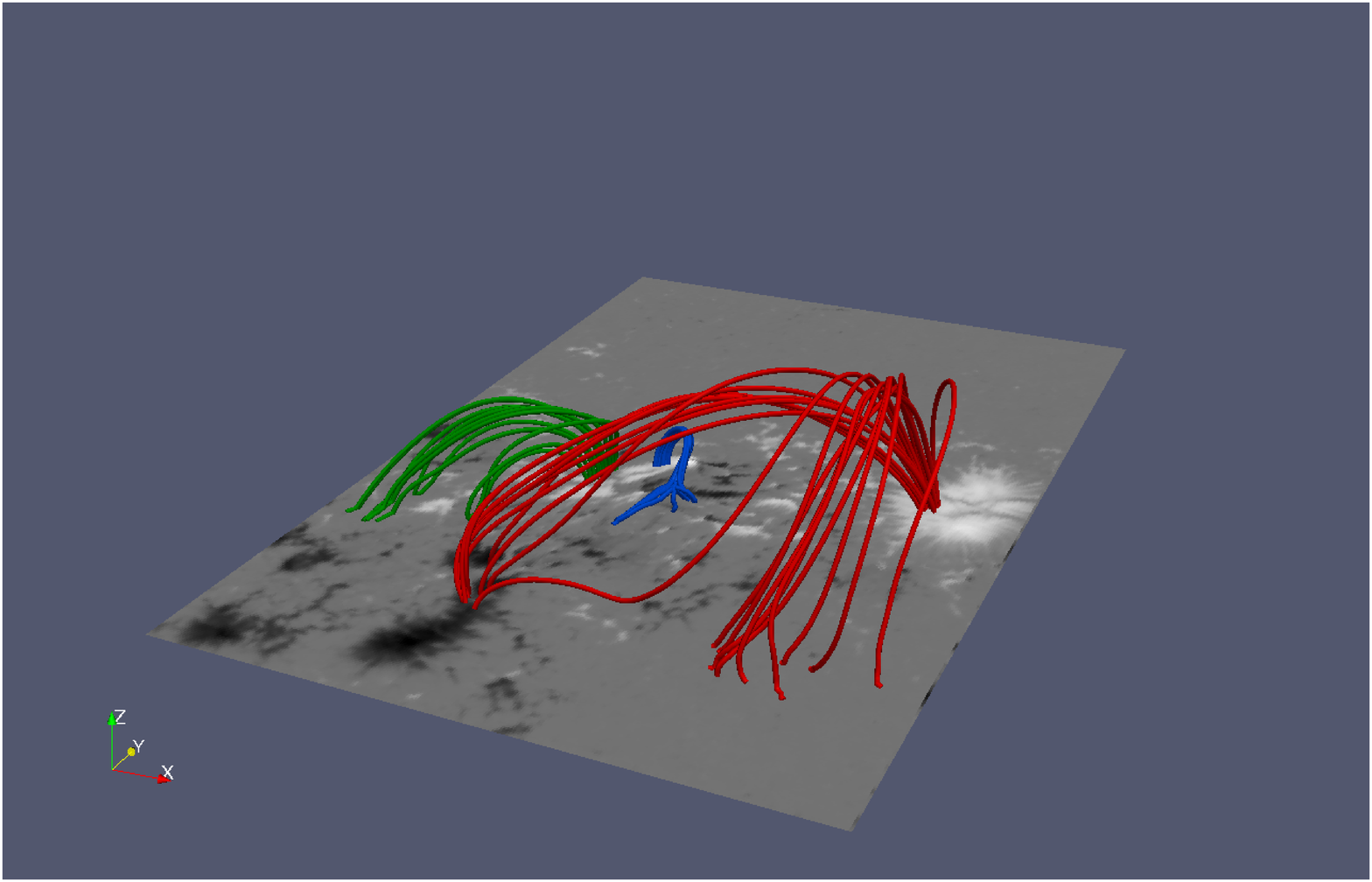}
                  }
 \raisebox{0pt}[0pt][0pt]{\raisebox{58mm}{\textcolor{white}{\textbf{
       \hspace{0mm} a) $\BDD$ \hspace{35mm} b) $\BTD$  \hspace{35mm} c) $\BMHD$  }}}}
 \raisebox{0pt}[0pt][0pt]{\raisebox{33mm}{\textcolor{white}{\textbf{
        \hspace{0mm} d) $\BEXsq$ \hspace{35mm} e) $\BEXfep$ \hspace{35mm} f) $\BEXfe$  }}}}
 \vspace{-8mm}
 \caption{Selected field lines of the six test cases:
(a) the potential field of a double dipole, $\BDD$;
(b) the TD model, $\BTD$;
(c) the MHD model, $\BMHD$;
(d) the NLFFF model of the nonpreprocessed magnetogram of AR~11158, $\BEXsq$;
(e) the NLFFF model of the preprocessed magnetogram of AR~11024, $\BEXfep$;
(f) the NLFFF model of the nonpreprocessed magnetogram of AR~11024, $\BEXfe$.
The vertical component of the magnetic field at the bottom boundary is shown on a gray scale, with the positive (respectively, negative) polarity in white (respectively, black).
The different line colors outline different types of connectivities.
}
 \label{f:fl}
\end{figure*}
To explore the effects of a finite divergence in a representative sample of practical  situations, we consider six test fields $\vBt$ obtained from analytical models, numerical simulations, and NLFFF extrapolations.
Their magnetic configuration is outlined in the field-line plots in \fig{fl}.
Furthermore, we consider six additional test cases $\vBts$, which are obtained from each of the $\vBt$ by removing the nonsolenoidal part of the field.

\subsection{Discretized analytical test fields}\label{s:testfields_a}
The first test field that we consider is the potential field $\vBt=\BDD$ generated by a pair of vertical magnetic dipoles, located at  $(0, \pm y_{DD}, z_{DD})$, see, \eg Eq.~(7) in \citet{2003A&A...406.1043T} for the analytical expression of the field.
We set $y_{DD}=2$ and $z_{DD}=-1.5$, and the field is normalized such that the $z$-component has a maximum value equal to unity at the bottom boundary ($z=0$).
The only currents and finite divergence errors present in $\BDD$ are generated by truncation errors in its discretization.

The second employed test field, $\vBt=\BTD$, is the model of the magnetic field of an active region derived in \cite{1999A&A...351..707T}, given by a section of a current ring surrounded by a stabilizing potential field.
The employed configuration is the same as in \citet{2012SoPh..278..347V}, to which we refer the reader for further details.
In this case, the test field has an explicit current-carrying component sustained by a flux rope.
The analytical formulae defining the test field are approximate, which together with the rather coarse resolution employed here, yield relatively large finite-divergence errors.

For both test fields  $\BDD$ and  $\BTD$ the discretized volume is $\vol=[-12,12] \times [-19, 19] \times [0, 16]$, with uniform resolution $\Delta=0.12$ in all directions.

\subsection{Numerical tests fields}\label{s:testfields_n}
The next test field that we consider, $\vBt=\BMHD$, is a snapshot of a magneto-hydrodynamic numerical simulation of magnetic reconnection in a null-point topology \citep{2012SoPh..276..199M}.
To use our present-stage diagnostic, we interpolated the original snapshot onto a uniform and homogeneous grid, whereas the original simulation was performed using a nonuniform one.
Because the divergence values are slightly increased by the interpolation, they are not representative of the quality of the simulations presented in \cite{2012SoPh..276..199M}.
However, they still serve our purpose of providing a typical situation arising from the numerical evolution of magneto-hydrodynamic equations.
The considered volume is  $\vol=[-20, 0] \times [-20, 10] \times [0, 12]$ with uniform resolution $\Delta=0.05$ in all directions, and the field is normalized such that the vertical component is unity at its maximum.

Next, we consider three NLFFF extrapolations of \textit{Hinode}/SOT vector magnetograms,  obtained with the magneto-frictional method in \citet{2010A&A...519A..44V}. 
The original resolution of the vector magnetograms is 0.3$''$, and they can be preprocessed \citep{2011A&A...526A..70F} to improve their compatibility with the force-free assumption on which the extrapolation code is based.

Our fourth test field,  $\vBt=\BEXsq$, is the nonlinear extrapolation of a vector magnetogram of AR~11158, measured on 14 February 2011.
The vector magnetogram was binned to the resolution $\Delta=1.1''$ prior to extrapolation, and  no preprocessing was applied in this case. 
The analyzed coronal model volume in arcsec is $\vol=[-21, 68] \times [-273, -171] \times [0, 123]$. 
The \textit{Hinode}/SOT field of view of the measurements employed for this extrapolation cuts through the external sunspots of a quadrupolar field distribution, resulting in high field values at the lateral edges of the magnetogram.
Even computing the potential field is problematic in this case, therefore we limited the considered volume to the bipolar core of the extrapolated field.  

The fifth test field, $\vBt=\BEXfep$, is the extrapolated coronal field model above AR~11024 on 4 July 2009.
In this case, the full resolution of \textit{Hinode}/SOT is used, and the vector magnetogram is preprocessed before extrapolation. 
The extrapolation covers a volume of $\vol=[-41, 42] \times [-141, -16] \times [0, 98]$ arcsec, with uniform resolution $\Delta=0.3''$.
This model of the coronal field of AR~11024 is discussed in detail in \citet{2011SoPh..tmp..374V}, where more details about extrapolation of vector magnetograms can be found.

Finally, the sixth test field, $\vBt=\BEXfe$, is the same case as  $\BEXfep$ except that the vector magnetogram is not preprocessed prior to extrapolation.
More details on the numerical implementation are given in \app{num_details}.
\subsection{Cleaned test fields} \label{s:cleanedfields}
Since a small divergence of $\vB$ is one major condition for the Thomson theorem, \eq{kelvin_exact}, for each test field $\vBt$ we consider a corresponding solenoidal version of it, $\vBts$, which is computed from $\vBt$ employing the divergence cleaner described in \app{cleaner}.
In Cartesian coordinates, such a solenoidal field has the same $x$- and $y$-components as  $\vBt$, whereas the $z$-component is changed everywhere in the volume, except for the top boundary.
Therefore, $\vBt$ and $\vBts$ have the same distribution of normal field on all boundaries except for the bottom one, where $\vBts$ differs from $\vBt$  by an amount that is related to the combined effect of $\divBt$ in the whole volume.
Since the divergence cleaner changes the value of the normal field component on one boundary, the potential fields computed from the boundary values of $\vBt$ and of the corresponding solenoidal $\vBts$ are not the same.
Additionally, the divergence cleaner alters the current of the field, as prescribed by \eq{jcleaner}, of an amount that is proportional to the divergence of $\vBt$.
Therefore, the field that is obtained by applying the cleaner may have drastically different properties than the original field. 
Finally, let us notice that different solenoidal fields can be derived from $\vBt$ using different methods.
The divergence-cleaned versions of the test fields $\vBts$ are used here as illustrative examples.

\section{Numerical tests of Thomson's theorem}\label{s:test_Thompson}
\begin{table*}
\caption{Numerical tests of Thomson's theorem.}
\label{t:thompson}
\strtable
\begin{tabular}{l c c | c c c c c |c}
\hline \hline
$\vBt$     & $\avfi{(\vB)}$  & $E$   & $\Epsn$  &$\EJsn$             &$\EdivBpn$        & $\EdivBJn$ & $\Emixn$ &Sum   \\
\hline
$\BDD$     & $ 2\times 10^{-6}  $& 1.45 & 1.00 & 0.00 &$ 4\times 10^{-5} $&$ 7\times 10^{-6}  $&$ -4\times 10^{-3} $& 1.00 \\
$\BTD$     & $ 3\times 10^{-6}  $& 3.90 & 0.81 & 0.16 &$ 3\times 10^{-5} $&$ 6\times 10^{-4}  $&$  0.02            $& 1.00 \\
$\BMHD$    & $ 2\times 10^{-5}  $& 1.94 & 0.94 & 0.06 &$ 1\times 10^{-6} $&$ 1\times 10^{-4}  $&$ -1\times 10^{-3} $& 1.00 \\
$\BEXsq$   & $ 4\times 10^{-3}  $& 4.21 & 0.79 & 0.38 &$ 5\times 10^{-4} $&$    0.29          $&$   -0.46          $& 1.00 \\
$\BEXfep$  & $ 9\times 10^{-4}  $& 1.51 & 0.88 & 0.11 &$ 2\times 10^{-4} $&$    0.14          $&$   -0.12          $& 1.00 \\
$\BEXfe$   & $ 2\times 10^{-3}  $& 0.72 & 2.29 & 0.14 &$ 3\times 10^{-4} $&$    0.94          $&$   -2.38          $& 1.00 \\
 \hline
$\vBts$    & & & & & & & &   \\
\hline                            
$\BDDs$    & $1\times 10^{-18} $& 1.44 & 1.00 & 0.00 &$ 4\times 10^{-5} $&$ 5 \times 10^{-5} $&$  7\times 10^{-4} $& 1.00 \\
$\BTDs$    & $4\times 10^{-21} $& 3.95 & 0.84 & 0.16 &$ 3\times 10^{-5} $&$ 4 \times 10^{-5} $&$ -3\times 10^{-4} $& 1.00 \\
$\BMHDs$   & $3\times 10^{-21} $& 1.94 & 0.94 & 0.06 &$ 1\times 10^{-6} $&$ 1 \times 10^{-6} $&$ -1\times 10^{-5} $& 1.00 \\
$\BEXsqs$  & $6\times 10^{-18} $& 5.98 & 0.43 & 0.57 &$ 2\times 10^{-4} $&$ 5 \times 10^{-3} $&$ -6\times 10^{-3} $& 1.00 \\
$\BEXfeps$ & $2\times 10^{-17} $& 3.15 & 0.42 & 0.58 &$ 1\times 10^{-4} $&$ 2 \times 10^{-3} $&$ -3\times 10^{-3} $& 1.00 \\
$\BEXfes$  & $8\times 10^{-18} $& 0.99 & 0.61 & 0.39 &$ 2\times 10^{-4} $&$ 1 \times 10^{-3} $&$ -2\times 10^{-3} $& 1.00 \\
\hline
\end{tabular}
\tablefoot{The employed test fields, defined in \sect{testfields}, are named in the leftmost column.
Second column, $\avfi{}$:  the divergence metric of the fields (see \eq{<fi>}).
Third column, $E$: energy of the test fields in units of $10^{32}$~erg. 
The $\BDD$, $\BTD$, $\BMHD$  fields (and their corresponding solenoidal fields  $\BDDs$, $\BTDs$, $\BMHDs$) were rescaled assuming a maximum value of the photospheric vertical field equal to 300~G and a typical distance between the sunspot's centers of (50, 50, 120)~Mm, respectively.   
The successive five columns are the different contributions  to \eq{kelvinn}, and ``Sum" corresponds to their sum.
All terms from ``$\Epsn$" to ``Sum" are normalized by $E$.
$\Epsn$ is the magnetic energy of the potential field $\vB_{\rm p,s}$,  $\EJsn$ that of the solenoidal component of the current-carrying one $\vB_{\rm J,s}$, $\EdivBpn$ and $\EdivBJn$ are the contributions associated to the divergence of $\vBp$ and $\vB_J$, respectively, $\Emixn$ is a mixed potential-current carrying term (see \eq{emixprime} for their definitions).
}
\end{table*}

In this section we apply \eq{kelvinn} to the test cases described in \sect{testfields}.
Table~\ref{t:thompson} summarizes  the values of the divergence metric $\avfi{}$ defined in \app{numdivb} and the contribution of each term to \eq{kelvinn}, for all test fields. 
The divergence metric spans values from $10^{-21}$ to $10^{-3}$. 
In all cases, the rightmost column, corresponding to the sum of the righthand side of \eq{kelvinn}, is equal to unity, despite the large difference in the divergence values. 
Therefore, we conclude that \eq{kelvinn} completely accounts  for all relevant contributions to the energy, in all test cases.
We then consider the different contributions to \eq{kelvinn} case by case.

\subsection{Results with the test fields}\label{s:ThompsonTest}
The top part of Table~\ref{t:thompson} refers to the test fields $\vBt$.
In general, the  energies $E$ of the different test fields go from the purely potential case of $\BDD$, where $\Epsn=1$, to high-free-energy cases ($\BTD$ and $\BEXsq$, with $\Epsn\simeq0.8$), where the field is strongly nonpotential. 
The main source of violation of Thomson's theorem, \eq{kelvin_exact}, in all cases is the mixed current-potential term $\Emixn$, except for the $\BEXfep$ case where $\EdivBJn$ is slightly higher in absolute value than $\Emixn$.

More precisely, $\BDD$ is nearly perfectly potential, with nonsolenoidal spurious fluctuations contributing to the total energy for few parts per thousand at most (in $\Emixn$).
$\BTD$ has a 16\%-energy contribution from the current-carrying part of the field $\EJsn$, with a 2\% contribution from the nonsolenoidal field related to the current-carrying structure (in $\Emixn$ but not in $\EdivBJn$). 
This is the effect of the approximate nature in the matching between current-carrying and external potential fields in the equilibrium that defines the $\BTD$ field.
$\BMHD$, which has 6\% free energy $\EJsn$, has an even lower nonsolenoidal contribution (-0.1\%).
In all three cases, there is very small ($\BTD$) or no significant ($\BDD$, $\BMHD$) violation of Thomson's theorem.

We now move to the NLFFF extrapolations. 
These show values of $\avfi{}$, which are two-to-three orders of magnitude greater than in the first three cases.
The contribution of the nonsolenoidal part of the potential field to the total energy, $\EdivBpn$, is always negligible with respect to the other terms.
In the $\BEXfep$ case, the free energy associated with the solenoidal part of the current-carrying field $\EJsn$ is about 11\%, and the potential field energy is 88\% of the total energy.
The sum of the potential and current-carrying solenoidal parts accounts for 99\% of the total energy, apparently verifying Thomson's theorem accurately.
However, $\EdivBJn$ is 14\% and $\Emixn$ is -12\%; \ie the errors related to the divergence of the current-carrying part of the field have comparable magnitudes and compensate for each other.
These are the dominant sources of error, almost three orders of magnitude more than  $\EdivBpn$.

The test case with the highest value of $\avfi{}$ is $\BEXsq$. 
With respect to the $\BEXfep$ case, $\BEXsq$ is characterized by three times higher free energy $\EJsn$, twice the error on current $\EdivBJn$, and almost a four times larger error on $\Emixn$. 
Again, the last two are largely compensating each other.
We conclude that the interpolation to one third of the resolution used for $\BEXsq$ is less efficient than preprocessing (used for $\BEXfep$) in eliminating the source of violation of Thomson's theorem.

This situation is even more extreme in the case of  the extrapolation of the nonpreprocessed, noninterpolated magnetogram $\BEXfe$.
Although this case has a value of the mean divergence $\avfi{}$ that is only a factor two higher than for $\BEXfep$, and not even the highest one, it shows the most pathological behavior: The potential field has an energy 2.29 times the energy of the test field, which is downright unphysical according to \eq{kelvin_exact}.
Such a high value is compensated for by an equally high value of $\Emixn$ (-2.38).
On the other hand, the current-carrying part of the field $\EJsn$ accounts for 14\% of the energy, but the associated error $\EdivBJn$ is more than six times larger. 
Such large errors are related to the high values of the divergence---in particular at the bottom boundary---and their actual values are very sensitive to the numerical details of the computation.
Additional analysis of  $\BEXfe$ and $\BEXfep$ is discussed in \sect{smallscales}

We finally notice  that in the preprocessed case $\BEXfep$, the error from $\EdivBJn$ or $\Emixn$ might be considered as still tolerable if compared with the total energy (errors on vector magnetograms are similar, after all), but it seriously compromises the reliability of the free energy estimation, each one being as high as $\EJsn$ itself.
  
\subsection{Results with the cleaned test fields}\label{s:ThompsonCleaned}

We now consider the bottom part of Table~\ref{t:thompson} for the solenoidal fields.
The values of the divergence are drastically reduced in all cases to $10^{-17}$ or less, which shows that the cleaner in \app{cleaner} is an effective---and fast---way of removing the nonsolenoidal component of a discretized magnetic field. 
For the purpose of this article, we can then consider all $\vBts$ to numerically be perfectly solenoidal.
All error terms, \ie $\EdivBpn$, $\EdivBJn$, and $\Emixn$, are smaller than 1\%, and we recover \eq{kelvin_exact} in a numerical sense.

More precisely,  the $\BDDs$ and $\BMHDs$ cases are practically identical to their corresponding test fields, as far as the energy metrics $E$, $\Epsn$, and $\EJsn$ are concerned. 
On the other hand, $\BTDs$ shows an increase of about 1.3\% of the total energy, $E$, as a result of the removal of the error in $\Emixn$  of $\BTD$. 
The error removal affects the potential field energy $\Ep$ more, which raises about 5\% with respect to the energy of the potential field in $\BTD$ (in nonnormalized values), as a consequence of the cleaner's modification of the bottom boundary.
In contrast, the relative contribution of the current-carrying part $\EJsn$ is unaffected by the cleaner.
It is true that $\avfi{}$ differs by 15 orders of magnitude between $\BTD$ and $\BTDs$, but it is significant anyway that the removal of a 2\%-error in $\Emixn$ changes the nonnormalized values of the total energy $E$ and potential field energy $\Eps$ of 1\% and 5\%, respectively.
We conclude that, even in relatively divergence-free fields, residual nonsolenoidal effects can be energetically significant.

In the extrapolated cases, the removal of the larger divergence has far stronger consequences.
In the first place, the nonnormalized field energy $E$ of the cleaned fields $\BEXsqs$, $\BEXfeps$, and $\BEXfes$ is increased of 42\%, 109\%, and 38\%, respectively, with respect to those of the corresponding test fields.
As a consequence of the higher values of $E$, the importance of potential fields relative to the total energy  $\Epsn$ is decreased (to 78\%, 95\%, and 40\% of their test-field values, respectively). 
In contrast, the energy contribution related to the current-carrying part of the field $\EJsn$ is strongly increased, as expected, since the cleaner introduces currents that are related to the cumulated divergence that is removed (see \app{cleaner}). 

We conclude that the cleaned fields that are obtained from the test ones using the method in \app{cleaner} all comply with Thomson's theorem accurately. 
However, three of them, namely $\BEXsqs$, $\BEXfeps$, and $\BEXfes$, are energetically very different from the original fields $\BEXsq$, $\BEXfep$, and $\BEXfe$, respectively. 
Incidentally, we notice that the removal of the finite divergence does not conserve the approximate force-freeness of the extrapolated fields.

\subsection{Contributions to $\Emixn$ for the test fields} \label{s:contrib_Emix}
\begin{table}
\caption{Contributions to $\Emixn$ in \eqs{emixprime}{kelvinn}.}
\label{t:emix_gi}
\centering
\strtable
\begin{tabular}{@{}l@{} c@{~~} c@{~~} c@{~~} c@{~~} c@{~~} c@{~~} |@{~} c@{}}
\hline \hline
$\vBt$     &  $\En_{\rm p,s / p,ns}$   &  $\En_{\rm J,s/J,ns}$ & $\En_{\rm p,s/J,ns}$ & $\En_{\rm J,s/p,ns}$ & $\En_{\rm p,ns/J,ns} $& $\En_{\rm p,s/J,s}$ &$ \Emixn$  \\
\hline
$\BDD$     & -0.01 & ~0.00 & ~0.00 & -0.00 & -0.00 & ~0.00 & -0.00 \\ 
$\BTD$     & -0.01 & ~0.00 & ~0.03 & ~0.00 & -0.00 & -0.00 & ~0.02 \\
$\BMHD$    & -0.00 & ~0.00 & -0.00 & -0.00 & -0.00 & -0.00 & -0.00 \\ 
$\BEXsq$   & -0.01 & -0.08 & -0.43 & ~0.00 & ~0.00 & ~0.05 & -0.46 \\ 
$\BEXfep$  & ~0.00 & -0.03 & -0.10 & ~0.00 & ~0.00 & ~0.01 & -0.12 \\ 
$\BEXfe$   & ~0.00 & -0.28 & -2.46 & ~0.00 & -0.00 & ~0.36 & -2.38 \\ 
\hline
\end{tabular}
\tablefoot{
$\En_{\rm p,s / p,ns} = \frac{1}{E}\intv \vB_{\rm p,s}\cdot \Nabla\zeta \ \dV$, \quad
$\En_{\rm J,s/J,ns}   = \frac{1}{E}\intv \vB_{\rm J,s}\cdot \Nabla\psi    \ \dV$,\\
$\En_{\rm p,s/J,ns}   = \frac{1}{E}\intv \vB_{\rm p,s}\cdot \Nabla\psi    \ \dV$, \quad
$\En_{\rm J,s/p,ns}   = \frac{1}{E}\intv \vB_{\rm J,s}\cdot \Nabla\zeta   \ \dV$,\\
$\En_{\rm p,ns/J,ns}  = \frac{1}{E}\intv \Nabla \zeta \cdot \Nabla\psi   \ \dV$, \quad
$\En_{\rm p,s/J,s}    = \frac{1}{E}\intv \vB_{\rm p,s}\cdot \vB_{\rm J,s} \ \dV$,\\
$\Emixn = \En_{\rm p,s / p,ns} +\En_{\rm J,s/J,ns}  +\En_{\rm p,s/J,ns} 
        +\En_{\rm J,s/p,ns}   +\En_{\rm p,ns/J,ns} +\En_{\rm p,s/J,s} $.
}
\end{table}
In many of the test fields in  Table~\ref{t:thompson}, $\Emixn$ is the largest source of error.
Table~\ref{t:emix_gi} shows the six contributions to $\Emixn$ in the order in which they appear in \eq{emixprime} and their sum $\Emixn$ for the six test cases  $\vBt$. 
We do not consider the solenoidal fields $\vBts$ since all terms are mostly zero and never bigger than 0.7\%
The following conclusions can be drawn. 
First, the main contribution to $\Emixn$ is $\En_{\rm p,s/J,ns}\equiv \frac{1}{E}\intv \vB_{\rm p,s}\cdot \Nabla\psi \ \dV$ in all cases. 
The main source of violation of Thomson's theorem is then the divergence of the current-carrying part of the field.
More often than not, this term has a similar magnitude and opposite sign of $\EdivBJn$, which is positive-definite. 
However, there is no obvious reason for $\En_{\rm p,s/J,ns}$ to be always---or predominantly--negative, and we regard this as a coincidence.

Second, the terms with residual divergence of the potential field (\ie any term containing $\Nabla \zeta$ in \eq{emixprime}) are always negligible. 
Therefore, also in view of the always low $\EdivBpn$ values in Table~\ref{t:thompson}, we can conclude that the divergence of the potential field always gives a negligible contribution to the energy.

Third, the integral $\En_{\rm J,s/J,ns}= \frac{1}{E}\intv \vB_{\rm J,s}\cdot \Nabla\psi \ \dV$, and the integral $\En_{\rm p,s/J,s}= \frac{1}{E}\intv \vB_{\rm p,s} \cdot \vB_{\rm J,s} \ \dV$ have finite values for the extrapolations in Table~\ref{t:emix_gi}. 
Analytically, they should be vanishing. Using the divergence theorem, \eq{divTheorem}, the surface integral vanishes because $\vB_{\rm J,s}|_\surf =0$, and the volume integral vanishes because $\Nabla \cdot \vB_{\rm J,s}=0$. 
The first condition is enforced at the boundary, but the second is only approximately true numerically (see also \sect{divergence}). 
This is not enough to insure that  $\En_{\rm J,s/J,ns}$ and $\En_{\rm p,s/J,s}$ vanish numerically.
This is why we adopted the decomposition of the energy of \sect{decomposition} that only contains volume integrals.

\section{Source of errors in the decomposition} \label{s:source_errors}
\subsection{Values of $\avfi{}$ for the field decomposition in  Eqs.~(\ref{eq:laplace}-\ref{eq:poisson})}\label{s:divergence}
\begin{table} 
\caption{Values of $\log_{10}(\avfi{})$, for the fields decomposition in Eqs.~(\ref{eq:laplace}), (\ref{eq:splitp}), and (\ref{eq:poisson}) (see \eq{<fi>} for the definition of $\avfi{}$).}
\label{t:fi}
\centering
\strtable
\begin{tabular}{@{~}l c@{\quad} c@{\quad} c@{\quad} c@{\quad} c@{\quad} c@{\quad} c@{~}}
\hline \hline
$\vBt$     &  $\vB$   & $\vBp$  & $\vB_{\rm p,s}$  & $\vB_{\rm p,ns}$   & $\vBJ$  & $\vB_{\rm J,s}$ &  $\vB_{\rm J,ns}$  \\
\hline 
$\BDD$     &  -5.61   &  -4.98  &   -5.60  &   -2.43   &  -2.35   &  -2.93   &  -2.46  \\  
$\BTD$     &  -5.54   &  -4.84  &   -5.40  &   -2.29   &  -4.22   &  -4.71   &  -3.20  \\
$\BMHD$    &  -4.78   &  -5.76  &   -6.23  &   -2.55   &  -3.96   &  -4.54   &  -2.45  \\ 
$\BEXsq$   &  -2.40   &  -2.60  &   -2.66  &   -1.42   &  -2.08   &  -2.36   &  -2.04  \\
$\BEXfep$  &  -3.05   &  -4.02  &   -4.09  &   -2.28   &  -2.62   &  -2.90   &  -2.62  \\ 
$\BEXfe$   &  -2.66   &  -3.87  &   -3.96  &   -2.12   &  -2.69   &  -2.86   &  -2.80  \\ 
 \hline
$\vBts$    & & & & & & &   \\
\hline
$\BDDs$    &  -18.0   &  -4.98  &   -5.60  &   -2.44   &  -0.83   &  -2.62   &  -1.91  \\ 
$\BTDs$    &  -20.4   &  -4.84  &   -5.40  &   -2.31   &  -3.41   &  -4.09   &  -2.15  \\
$\BMHDs$   &  -20.5   &  -5.42  &   -5.79  &   -2.39   &  -4.39   &  -4.70   &  -2.31  \\
$\BEXsqs$  &  -17.2   &  -2.65  &   -2.71  &   -1.47   &  -1.72   &  -2.11   &  -1.83  \\ 
$\BEXfeps$ &  -16.8   &  -3.78  &   -3.93  &   -2.26   &  -0.44   &  -2.76   &  -2.25  \\
$\BEXfes$  &  -17.1   &  -3.63  &   -3.79  &   -1.99   &  -1.51   &  -2.71   &  -2.19  \\
\hline
\end{tabular}
\tablefoot{Column $\vB $ here is the logarithm of the column $\avfi{(\vB)}$ in Table~\ref{t:thompson}.
More negative values correspond to more solenoidal fields.
}
\end{table}

In this section we quantify how accurate the decomposition in solenoidal and nonsolenoidal contributions is.
Table~\ref{t:fi} reports the values of the logarithm of $\avfi{}$, defined by  \eq{<fi>},  for the field decomposition used in \eq{kelvin}.
Since $\avfi{}$ is not additive in the field, its value for, say, $\vB$ is different from the sum of its values for the potential $\vBp$ and current-carrying $\vBJ$ components. 

We next consider the decomposition of the potential field given by \eq{splitp} for the test fields $\vBt$ (upper half of Table~\ref{t:fi}). 
Values of $\avfi{}$ for the solenoidal part of the potential field $\vBps$ are better (\ie more negative) than those for $\vBpns$, so that the $\vBps$ is indeed more solenoidal than $\vBpns$.
However, it is only in the first three cases, $\BDD$, $\BTD$, and $\BMHD$, that $\log_{10}(\avfi{})$ has a noticeably  more negative value for $\vBps$ than for $\vBp$. 
In the other cases, the values are relatively close to each other, and  $\vBps$ is only marginally more solenoidal than $\vBp$. 
On the other hand, $\vBpns$ is always  much less solenoidal than  $\vBp$.
This is partly the effect of the nonadditivity of the metric $\avfi{}$, and partly because $\vBpns$ is, on average, much smaller than $\vBps$, as the corresponding energy metrics in  Table~\ref{t:thompson} show. (In particular, $\EdivBpn$, \ie the energy associated with $\vBpns$, is always extremely small.)

Similar conclusions can be drawn looking at the decomposition of the current-carrying part, $\vBJ$, where this time the energy associated with the solenoidal error (see $\EdivBJn$ in Table~\ref{t:thompson}) is more significant.
In this case, values of  $\avfi{}$ for all three contributions $\vBJ$, $\vBJs$, and $\vBJns$ are of similar magnitude.
Again, the nonsolenoidal part, $\vBJns$, has a higher divergence value than the solenoidal one, $\vBJs$, but only marginally so for $\BDD$ and extrapolated fields.

We consider the solenoidal test fields $\vBts$ (bottom half of Table~\ref{t:fi}).
Values of $\avfi{}$ for a given field component belonging to $\vBt$ and to the corresponding  $\vBts$ are very similar.
For instance, the value of  $\log_{10}(\avfi{})$ for $\vB_{\rm J,s}$ in, say, the test field $\BEXsq$ is $-2.36$, whereas for the corresponding contribution for $\BEXsqs$ it is $-2.11$. 
Therefore, the above discussion of the contributions to $\vBt$ holds for those of $\vBts$ as well.
In contrast, the total divergence of the field is very different in the two cases, \ie -2.4 and -17.2, respectively. 
This is a clear indication that the accuracy of the field decomposition is determined by the accuracy in the solution of Eqs.~(\ref{eq:laplace}-\ref{eq:poisson}) rather than by the divergence of the total field.

In conclusion, the Poisson solver provides a decomposition of the magnetic field where the solenoidal parts have a smaller divergence than the original field, as required. 
The limit in the accuracy of the decomposition comes from the accuracy of the solver, and not from the level of solenoidality of the initial field.
One possible source of inaccuracy for the solver is the incompatibility of the boundary conditions used in  Eqs.~(\ref{eq:laplace}-\ref{eq:poisson}), which is  discussed in the next section.

\subsection{Compatibility of boundary conditions in Eqs.~(\ref{eq:laplace}-\ref{eq:poisson}).}
\label{s:imbalance}
\begin{table}
\caption{Removal of flux imbalance.}
\label{t:fluxbal}
\centering
\strtable
\begin{tabular}{@{~}l@{~} c c@{\quad}  c@{\quad}  c@{\quad}  c@{~} }
\hline \hline
$\vBt$     &  $\avfi{(\vBt^{\rm bal})}$&$\Phi_\surf(\vBt)$&$\Phi_\surf(\vBt^{\rm bal})$&$\En_{\Phi}$&$\En_{\Phi,\rm mix}$  \\
\hline
$\BDD$     &  -5.61  &   -6.83  &  -26.4  &  -17.3   & -11.2 \\ 
$\BTD$     &  -5.54  &   -7.79  &  -26.0  &  -19.1   & -11.8 \\
$\BMHD$    &  -4.78  &   -4.36  &  -22.8  &  -12.2   & -6.29 \\ 
$\BEXsq$   &  -2.40  &   -1.33  &  -18.7  &  -5.18   & -3.30 \\ 
$\BEXfep$  &  -2.60  &   -2.00  &  -20.9  &  -7.49   & -4.52 \\ 
$\BEXfe$   &  -2.63  &   -1.86  &  -20.0  &  -6.77   & -4.35 \\ 
\hline
\end{tabular}
\tablefoot{For all quantities, the $\log_{10}$ of the absolute value is shown. $\avfi{}$ is the divergence metric defined in \eq{<fi>}; $\Phi_\surf$ is the normalized magnetic flux through all boundaries, \eq{fluxes_S}; $\En_{\Phi}$ and $\En_{\Phi,\rm mix}$ are normalized to the energy $E$ of $\vBt^{\rm bal}$, which is the flux balanced field associated to $\vBt$, see \eq{Bbal}.}
\end{table}
We here consider the  normalized flux of the field, $\Ps$, computed as the surface flux through all six boundaries, normalized to  the mean flux entering and exiting from the lower boundary:
 \BA 
  \Ps(\vB)           &=& \ints \vB \cdot \dS ~/ \Phi_{\rm norm} \,,   \label{eq:fluxes_S}\\
  {\rm with~}  \Phi_{\rm norm} &=&  \frac{1}{2} \int_{z=z_1} |\vB \cdot \dS| \,. \nonumber
  \EA  
The values of  $\log_{10}|\Phi_\surf(\vBt)|$ in Table~\ref{t:fluxbal} show that the test fields of the extrapolation cases $\BEXsq$, $\BEXfep$, and $\BEXfe$ are not flux-balanced. 
Therefore, the decomposition of \eq{kelvin} based on the solutions of Eqs.~(\ref{eq:laplace}-\ref{eq:poisson}) may be inconsistent (see \eq{gen_poisson} and related text).
The purpose here is to determine whether the unbalanced flux affects the accuracy of any of the terms in \eq{kelvin}.  

A flux-balanced field, $\vBt^{\rm bal}$, can be computed from a flux-unbalanced one, $\vBt$,  by splitting the original field as 
\BE
\vBt^{\rm bal}= \vBt + \vB_{\rm \Phi} \,, 
\label{eq:Bbal}
\EE
and assuming $\vB_{\rm \Phi}=\Nabla \Theta$ to be generated by an uniformly distributed, constant divergence; \ie $\Delta \Theta=$ constant. 
We choose the simple solution $\Theta \propto \vec{r}^2$, and fix the constant such that the flux of $\vB_{\rm \Phi}$ equals the flux of  $\vBt$, yielding
\BE  
\vB_{\rm \Phi}=\left ( \frac{1}{3 \mathcal{V}}\ints \vBt \cdot \dS \ \right) \vec{r}.
\nonumber
\EE   

Table~\ref{t:fluxbal} shows that the above modifications to $\vBt$ is effective, drastically reducing the net flux of the original field  to very low values (compare $\log_{10}|\Phi_\surf(\vBt)|$ with $\log_{10}|\Phi_\surf(\vBt^{\rm bal})|$). 
On the other hand, the effect on the field of $\vB_{\rm \Phi}$ is very small. 
Both energy terms related to that (\ie  $\En_{\Phi}\equiv\frac{1}{2E}\intv B^2_{\Phi} \dV$ and $\En_{\Phi,\rm mix}\equiv \frac{1}{E}\intv \vBt \cdot \vB_{\rm \Phi} \dV$) are negligible (with a contribution below 0.01\% of the total energy $E$ computed for $\vBt^{\rm bal}$).

Repeating the same analysis of Sects.~\ref{s:test_Thompson} and \ref{s:divergence} for the flux-balanced part of the field only, $\vBt^{\rm bal}$, yields no significant change: all values in Tables~\ref{t:thompson}, \ref{t:emix_gi}, and \ref{t:fi} are identical.
Inaccuracies of the Poisson solver in solving \eq{laplace} are therefore related to the solver itself, not to the incompatibility of the boundary conditions.

In a similar way, the test field can be modified to have vanishing volume divergence, which is the requirement for consistency in solving \eqs{splitp}{poisson}, using
\BE   
\vB_{\rm  \Phi}=\left ( \frac{1}{3 \mathcal{V}}\intv \Nabla \cdot \vBt \dV  \ \right) \vec{r} \,.
\nonumber \label{eq:BdPhiDiv}
\EE   
The result is likewise clear: no significant change is found in the values of Tables~\ref{t:thompson}, \ref{t:emix_gi}, and \ref{t:fi}.

Therefore, an imperfect consistency of source and boundary conditions play no role in the accuracy of the solution of the Laplace and Poisson equations employed in the decomposition, \eq{kelvin}, for any of the test cases. 
Recalling the results of \sect{divergence}, we conclude that the accuracy limitation of our analysis comes from the solver itself.
In this respect, we note that, when the  method used in Eqs.~(\ref{eq:splitp}) and (\ref{eq:poisson}) is viewed as an algorithm for removing the divergence (Projection method), it  is far less efficient than our divergence cleaner described in \app{cleaner}. 
On the other hand, the projection method has other advantages; for instance, it change neither the current in the volume nor the normal component of the original field at the boundaries.

\section{Parametric study}\label{s:Parametric}

In this section we study how the relative energy of the field depends on its divergence in progressively going from a solenoidal to nonsolenoidal realizations. 
The purpose is to offer a practical method of fixing the level of solenoidal errors that can be tolerated in a given numerical realization, based on their consequences on the energy of the field.

\subsection{Parametric models of finite-divergence fields.}\label{s:div-model}
 
For a given test magnetic field $\vBt$, the corresponding  solenoidal field $\vBts$ is considered. 
A parametric, nonsolenoidal field $\vBns$ is obtained by adding a nonsolenoidal component $\vBdiv$ to $\vBts$, using a control parameter $\delta$, as
  \BE 
  \vBns= \vBts +\delta ~ \vBdiv \, .
  \label{eq:btest}
  \EE
We consider here two models of  $\vBdiv$, namely
  \BE 
  \vBdiv = \left \{ 
    \begin{array}{rr} 
    -\hatz \int_{z}^{z_2} (\divBt) \rmd z' \qquad \qquad \;\;          & \mbox{Model 1} , \\
     -\frac{1}{3}\left ( \hat{\vec{x}}\int_{x}^{x_2} \rmd x' 
                    + \hat{\vec{y}}\int_{y}^{y_2} \rmd y' + \right. \qquad \;\;\,  &\\
               \left.  + \hatz \int_{z}^{z_2} \rmd z' 
                \right) \divBt            & \mbox{Model 2} .
    \end{array}
  \right. 
  \label{eq:divmodels}
  \EE
Adding the first divergence model for $\delta=1$ is the inverse operation of the cleaner in Section~\ref{s:cleaner}, since $\vBns(\delta=1)=\vBt$.  
For other $\delta$ values, the resulting field $\vBns$ only differs from the solenoidal field $\vBts$ in the $z$-component.
The second model for the divergence is more general, because it changes all three components of $\vBts$ in the volume, although not on all boundaries.

Both divergence models in \eq{divmodels} are based on the computed $\divBt$.
In this way, we relate the divergence models to the source of error that is specific to the considered test field.
For instance, we expect that errors in the  test case $\BDD$, which are only generated by truncation errors, have a different distribution in space than those coming from the approximate nature of the $\BTD$ equilibrium, or from a numerically constructed field like $\BEXfep$.

The influence of a finite divergence of $\vBns$ on the energy value can be written as
  \BE  
  E=\delta^2  \Ens+2\delta  \Esns +\Es \, ,
  \label{eq:enns}
  \EE
where $E$, $\Ens$, and $\Es$ are defined as usual as proportional to the volume integrals of $B^2_\delta$, $B_{\rm div}^2$, and  $B_{\rm test,s}^2$, respectively, and $\Esns\equiv \intv (\vBts \cdot \vBdiv) \ \dV/8\pi$. 

Below, the energy dependence on $\delta$ is studied for the test fields described in \sect{testfields}. 
Since the separation in solenoidal and nonsolenoidal components is known by construction, we simplify the presentation by analyzing the energies of the total fields according to \eq{enns}, and we do not separate the error sources as in \sect{test_Thompson}.
For each value of $\delta$, we consider $\vBns$ as the test field to analyze and compute the corresponding potential field according to \eq{laplace}.

 \begin{figure*}
 \centerline{
   \includegraphics[width=0.33\textwidth]{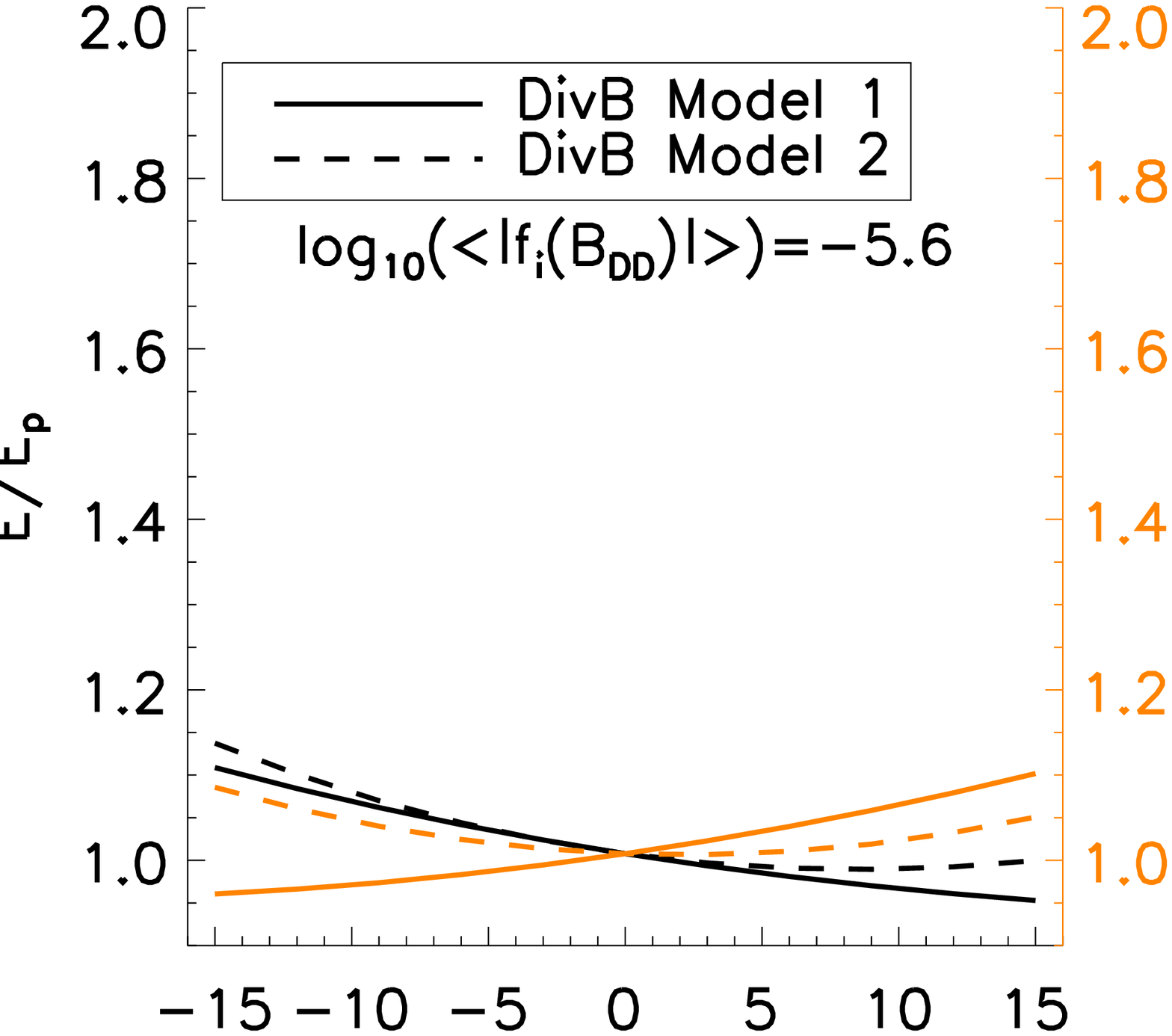}
   \includegraphics[width=0.33\textwidth]{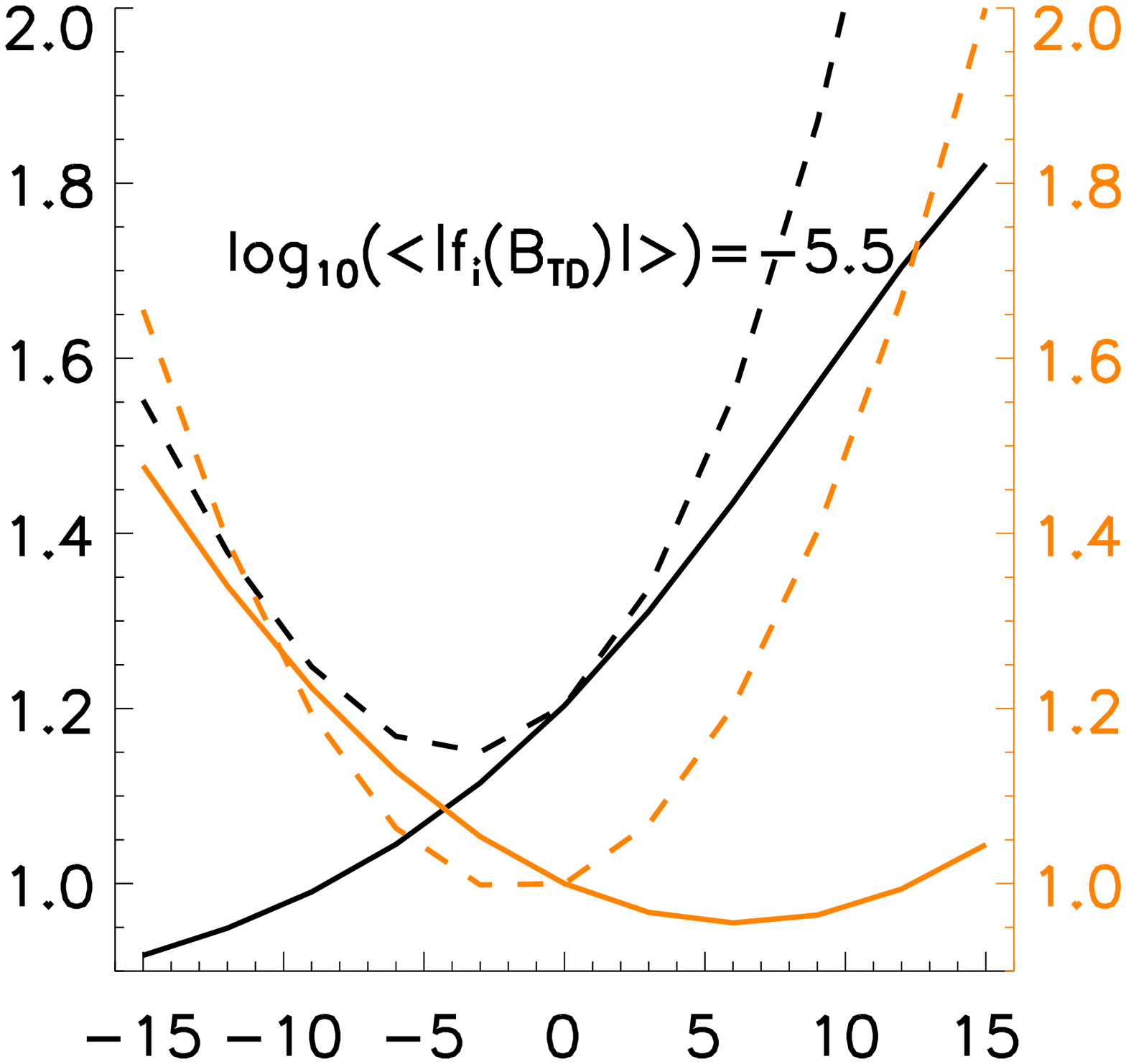}
   \includegraphics[width=0.33\textwidth]{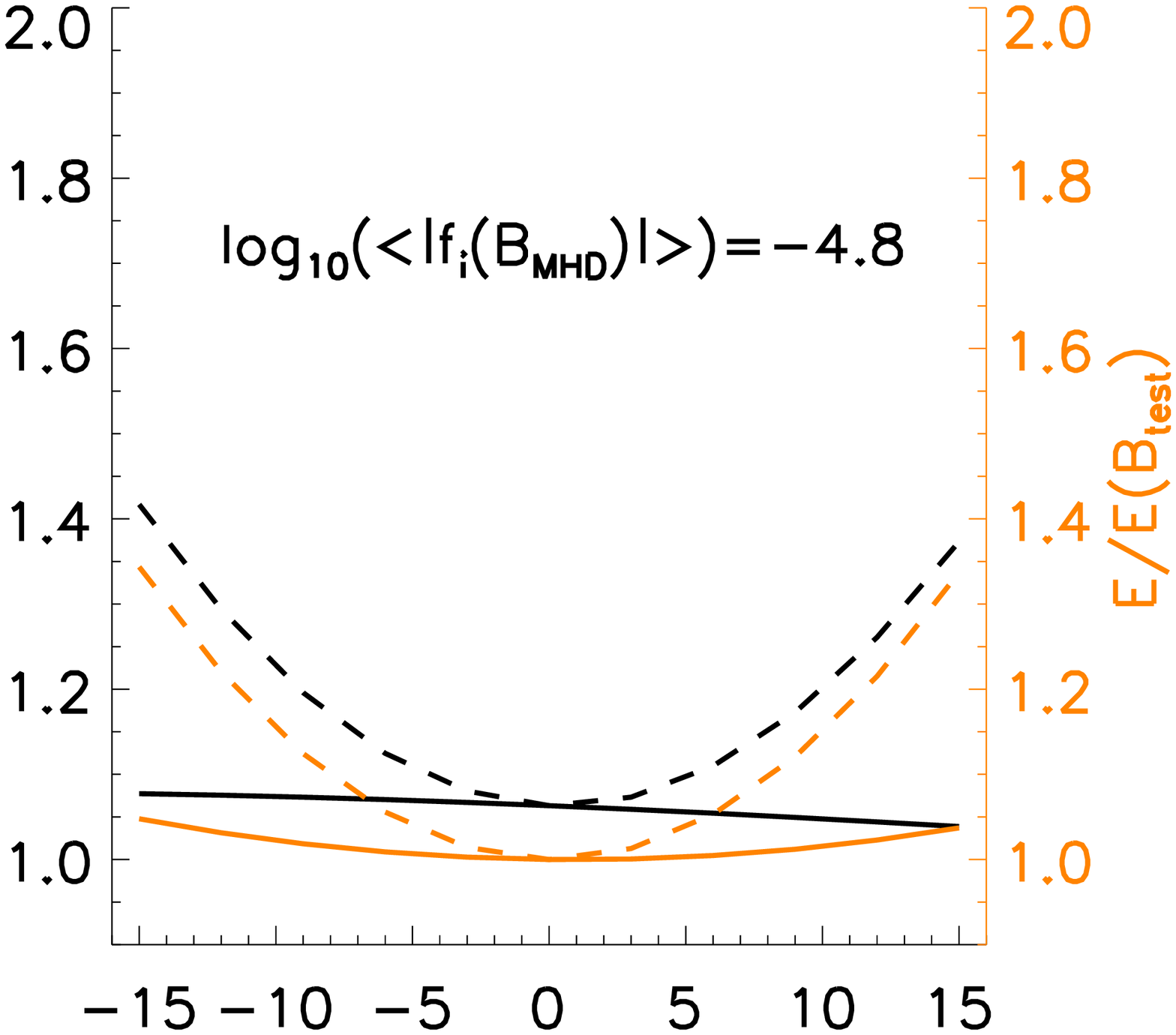}
                }
 \vspace{-4mm}
  \centerline{
   \includegraphics[width=0.33\textwidth]{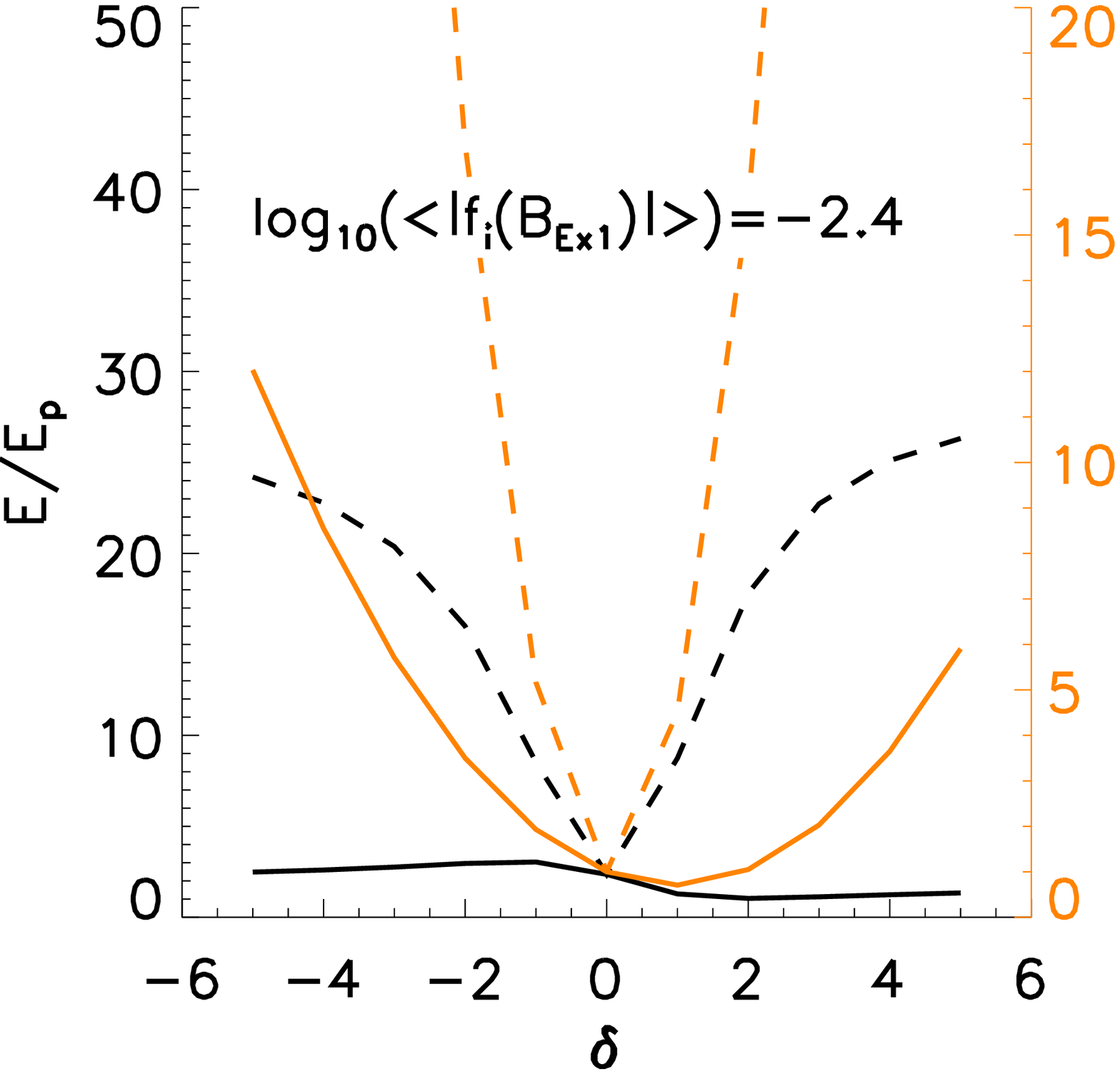}
   \includegraphics[width=0.33\textwidth]{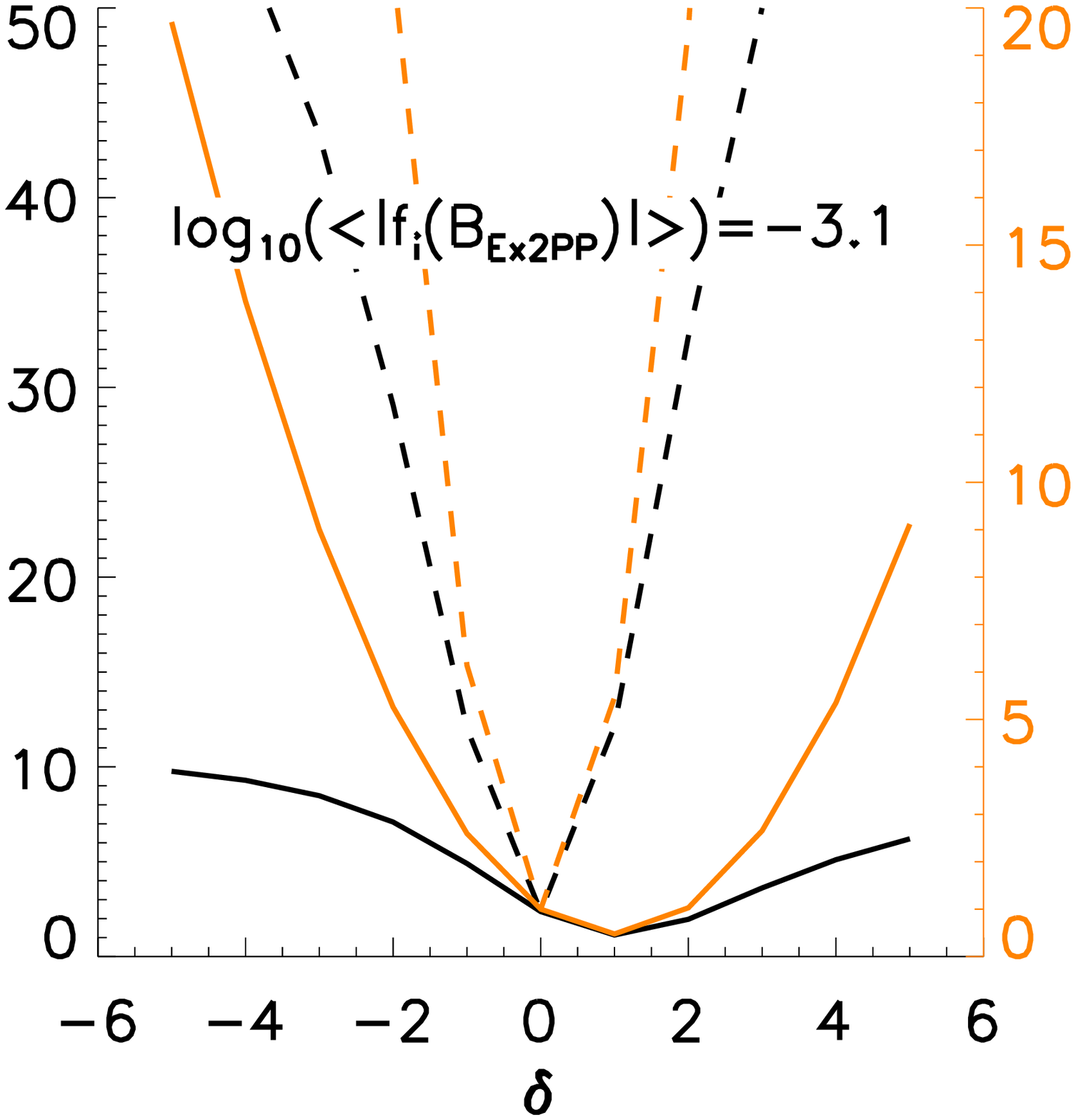}
   \includegraphics[width=0.33\textwidth]{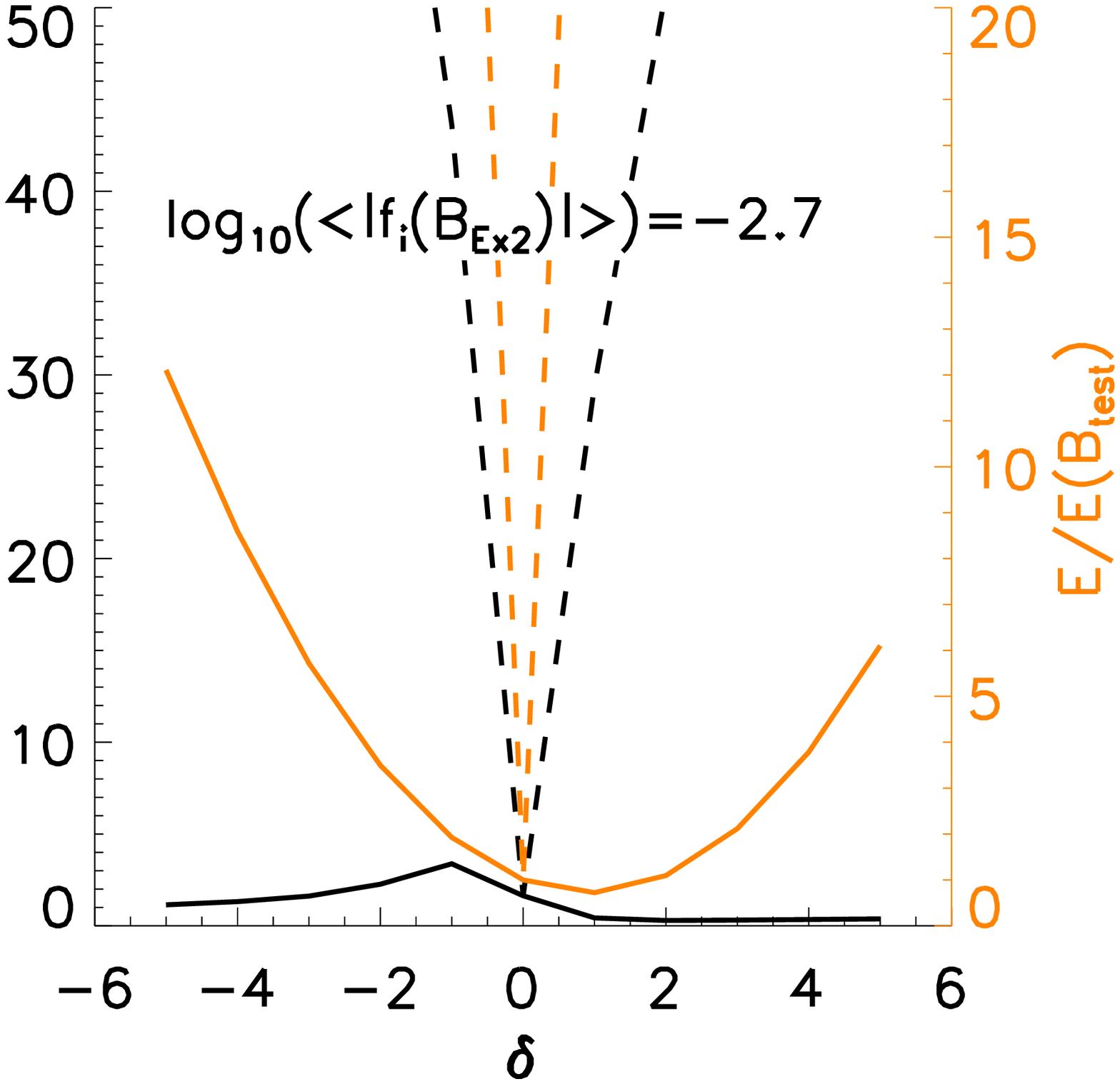}
                  }
 \raisebox{0pt}[0pt][0pt]{\raisebox{87mm}{\textcolor{black}{\textbf{
       \hspace{6mm} a) $\BDD$ \hspace{35mm} b) $\BTD$  \hspace{35mm} c) $\BMHD$  }}}}
 \raisebox{0pt}[0pt][0pt]{\raisebox{50mm}{\textcolor{black}{\textbf{
        \hspace{6mm} d) $\BEXsq$ \hspace{35mm} e) $\BEXfep$ \hspace{32mm} f) $\BEXfe$  }}}}
 \vspace{-8mm}
 \caption{Magnetic energy, \eq{enns}, normalized to the energy of the potential field having the same distribution of normal field on the boundaries (black lines) or to the energy of the reference field (orange lines), as a function of the amplitude $\delta$ of the nonsolenoidal term, \eq{btest}.
Two models for the divergence are shown according to \eq{divmodels}.  Each panel shows the results of a test field:
(a) the potential field of a double dipole, $\BDD$;
(b) the TD model, $\BTD$;
(c) the MHD model, $\BMHD$;
(d) the NLFFF model of the nonpreprocessed magnetogram of AR~11158, $\BEXsq$;
(e) the NLFFF model of the preprocessed magnetogram of AR~11024, $\BEXfep$;
(f) the NLFFF model of the nonpreprocessed magnetogram of AR~11024, $\BEXfe$.
An important change of scale of both axes is present between the top and the bottom rows.
}
 \label{f:endivb_sr}
\end{figure*}

\subsection{Parametric dependence of the energy}\label{s:par-dependance}
Figure~\ref{f:endivb_sr} shows the energy for the two divergence models in \eq{divmodels}, as a function of the control parameter $\delta$ in a wide range of values.
Due to the large difference in $\divBt$ between the six models, the top- and bottom-rows have different scales.  
The orange lines show the energy normalized with the energy of $\vBt$, which is not dependent on $\delta$: They
follow the expected parabolic profile of \eq{enns}, only scaled by the normalization factor. 
Model~1 (continuous orange lines) yields a smaller variation of the energy with $\delta$ (corresponding to lower values of $\Ens$) with respect to Model~2 (dashed orange lines), and is centered farther away from $\delta=0$ (\ie Model~1 has higher values of  $\Esns / \Ens$).

The orange curves in the top row of Figure~\ref{f:endivb_sr} show that it takes very high values of $\delta$ in order to have a variation of order one of the energy in the $\BDD$, $\BTD$, and $\BMHD$ cases (\eg for the Model~2 applied to $\BTD$ at $\delta=15$).
On the other hand, the energy of the extrapolated fields shows a much steeper increase with $\delta$, related to the much higher value of  $\divBt$, and particularly so for Model~2.

The location of the minimum of each of the orange curves is at $\delta_{\rm min}=-E_{\rm mix}/E_{\rm div}$, therefore its location depends on the average orientation and amplitude of the divergence field $\vBdiv$ with respect to the solenoidal field $\vBs$.
The orientation and amplitude of $\vBdiv$ also determines the height of the minimum (since the energy of the test field is fixed). 
With both divergence models, there are no general rules; \ie the energy can increase or decrease with $\delta$, and the location of the minimum depends on the case.
\subsection{Comparison with the potential field energy}\label{s:par-potential}
The physically meaningful quantity is represented by the energy normalized to the energy of the corresponding potential field, represented in Figure~\ref{f:endivb_sr} by black lines. 
For different $\delta$ values, the normal component of the field $\vBdiv$ at the boundary changes according to \eq{divmodels}, hence also the energy of corresponding potential field depends---quadratically---on $\delta$.  
Due to the additional $\delta$-dependence, the shape of the black lines is not always parabolic in the six cases, and the actual profiles depend on the details of the spatial distribution of divergence in the test field.

To show that, we first notice that the two divergence models behave very differently, except for $\BDD$ where the range in $\delta$ is too narrow to show significant differences.
For instance,  $E/ \Ep$ of Model~1 (continuous black lines) is an increasing function of $\delta$ in the range (-15, 15) in the $\BTD$ case. 
Model~2, on the other hand, has a parabolic energy profile with minimum at $\delta \approx -4$. 
For both models, the energy variation is relatively large (1.8 and above 2 for Models~1 and 2, respectively), whereas the variation in the same range of $\delta$ is smaller for the  $\BDD$ and $\BMHD$ cases.

The extrapolated cases yield not only much larger variations (note again the difference in scales  between the top and bottom rows of Figure~\ref{f:endivb_sr}), but also a stronger dependence on $\delta$.
In particular, Model~2 yields a relative energy that sharply increases with $\delta$, for instance, to one order of magnitude increase for $\delta$ going from the value 0 to 1 in the $\BEXfep$ case.
A saturation at high  values of $\delta$ is clearly visible in the dashed black line (Model~2) of the $\BEXsq$ case, and is hinted at in the $\BEXfep$ case.
Such saturation is actually present in all three extrapolated cases, yielding values that are higher than those shown in the corresponding plots.
The saturation happens when the quadratic dependence on $\delta$ of the energy of potential field compensates the quadratic term $\delta^2 \Ens$. 

\begin{figure*}
 \centerline{
   \includegraphics[width=0.33\textwidth]{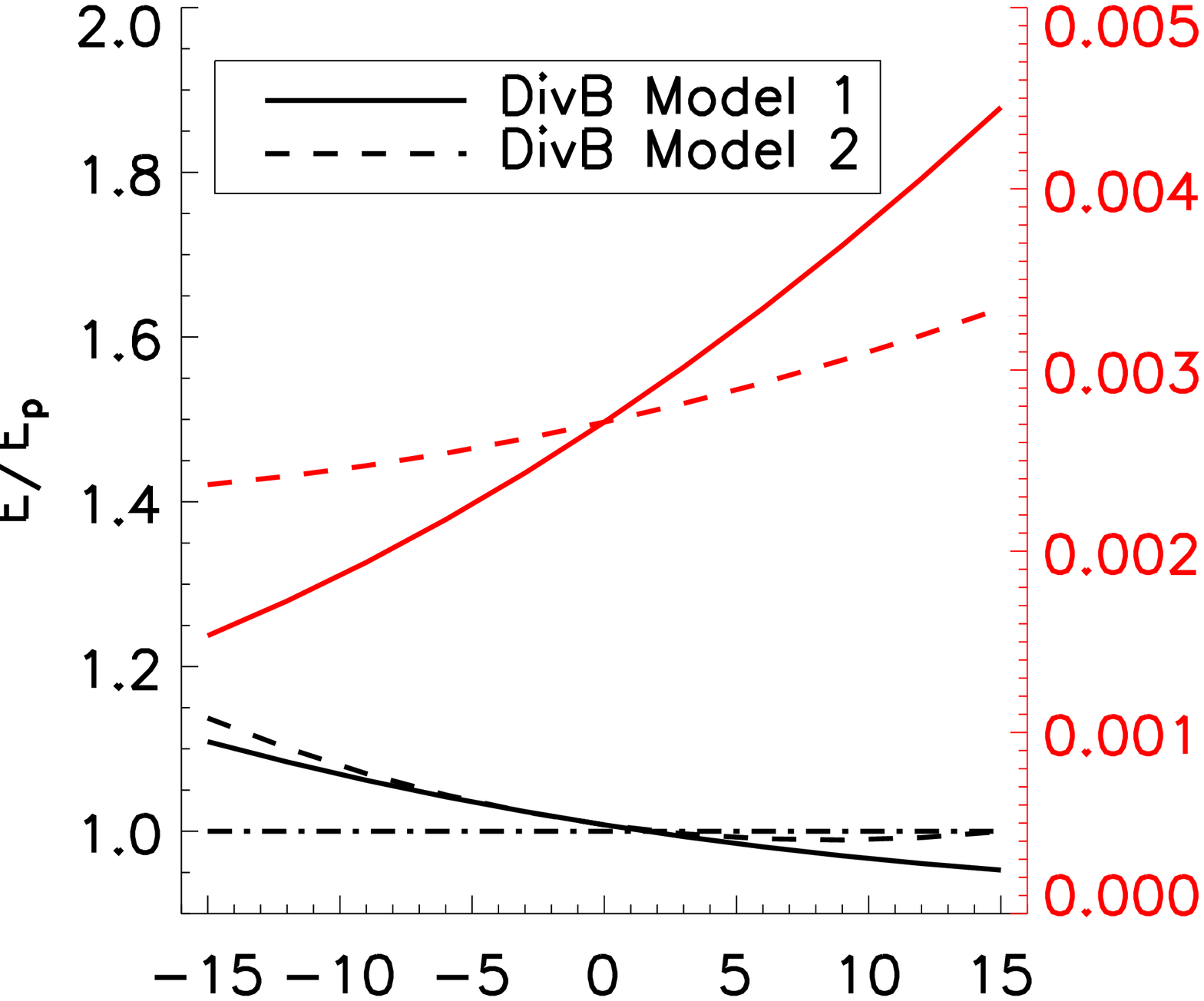}
   \includegraphics[width=0.33\textwidth]{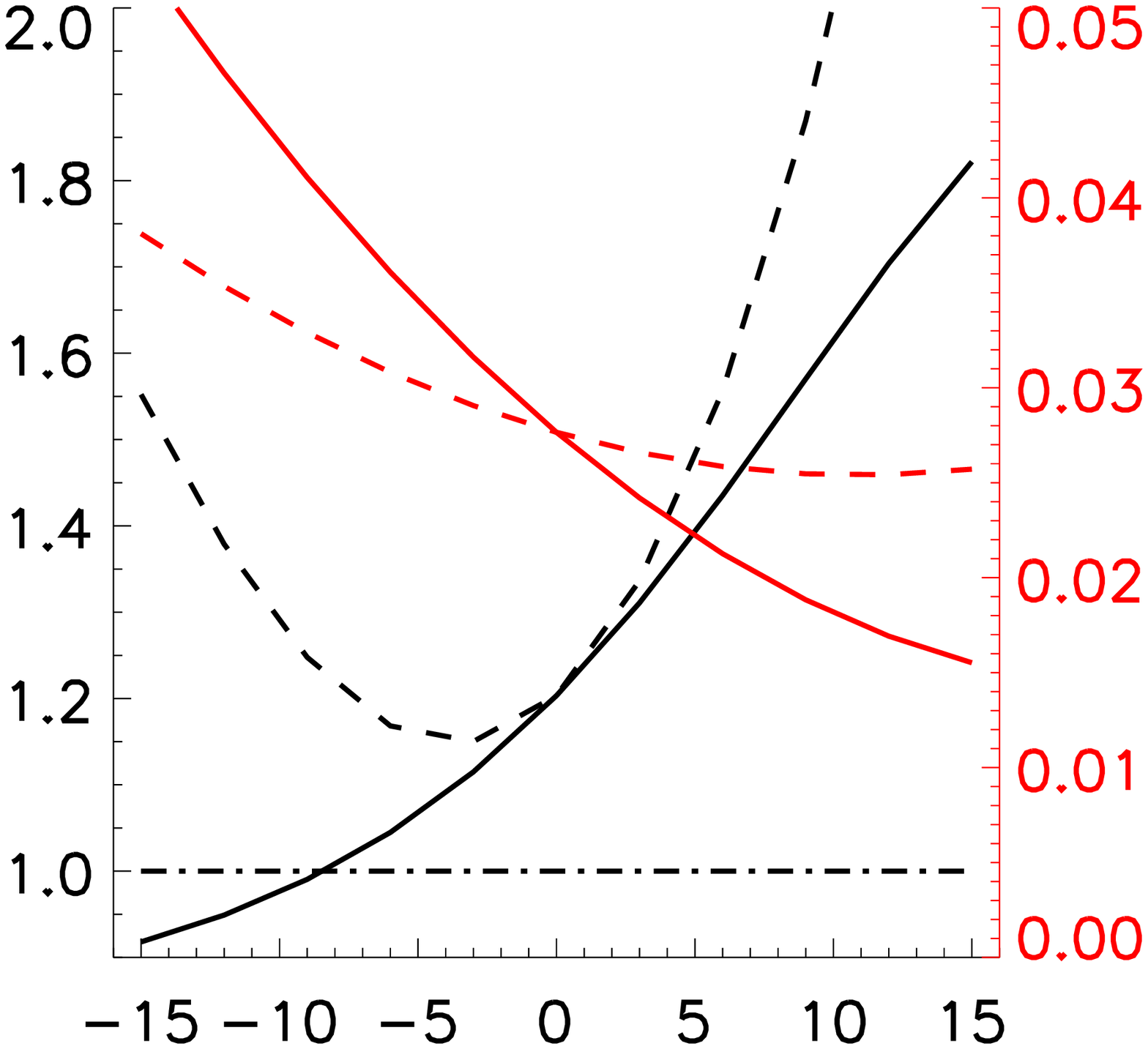}
   \includegraphics[width=0.33\textwidth]{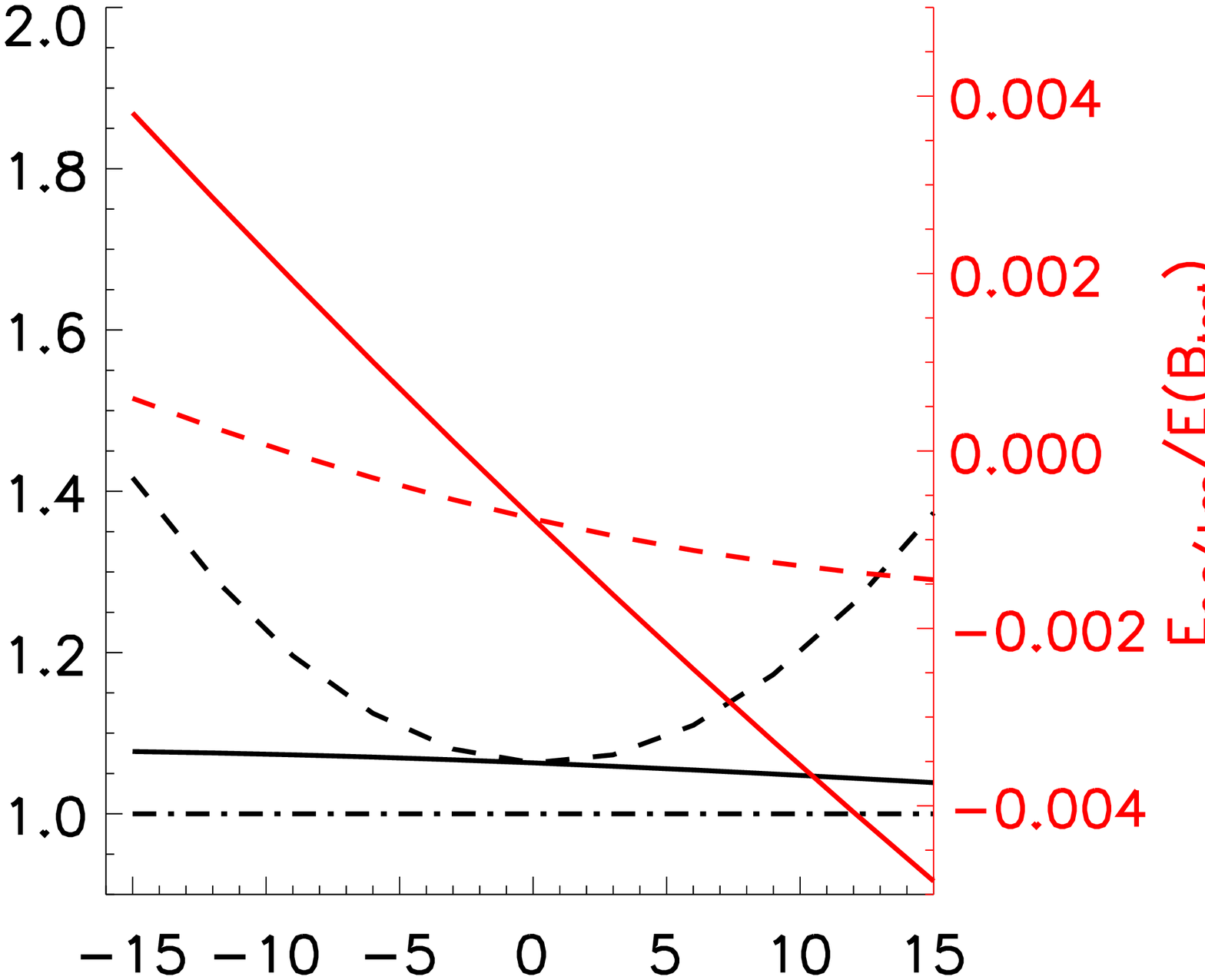}
                }
 \vspace{-4mm}
  \centerline{
   \includegraphics[width=0.33\textwidth]{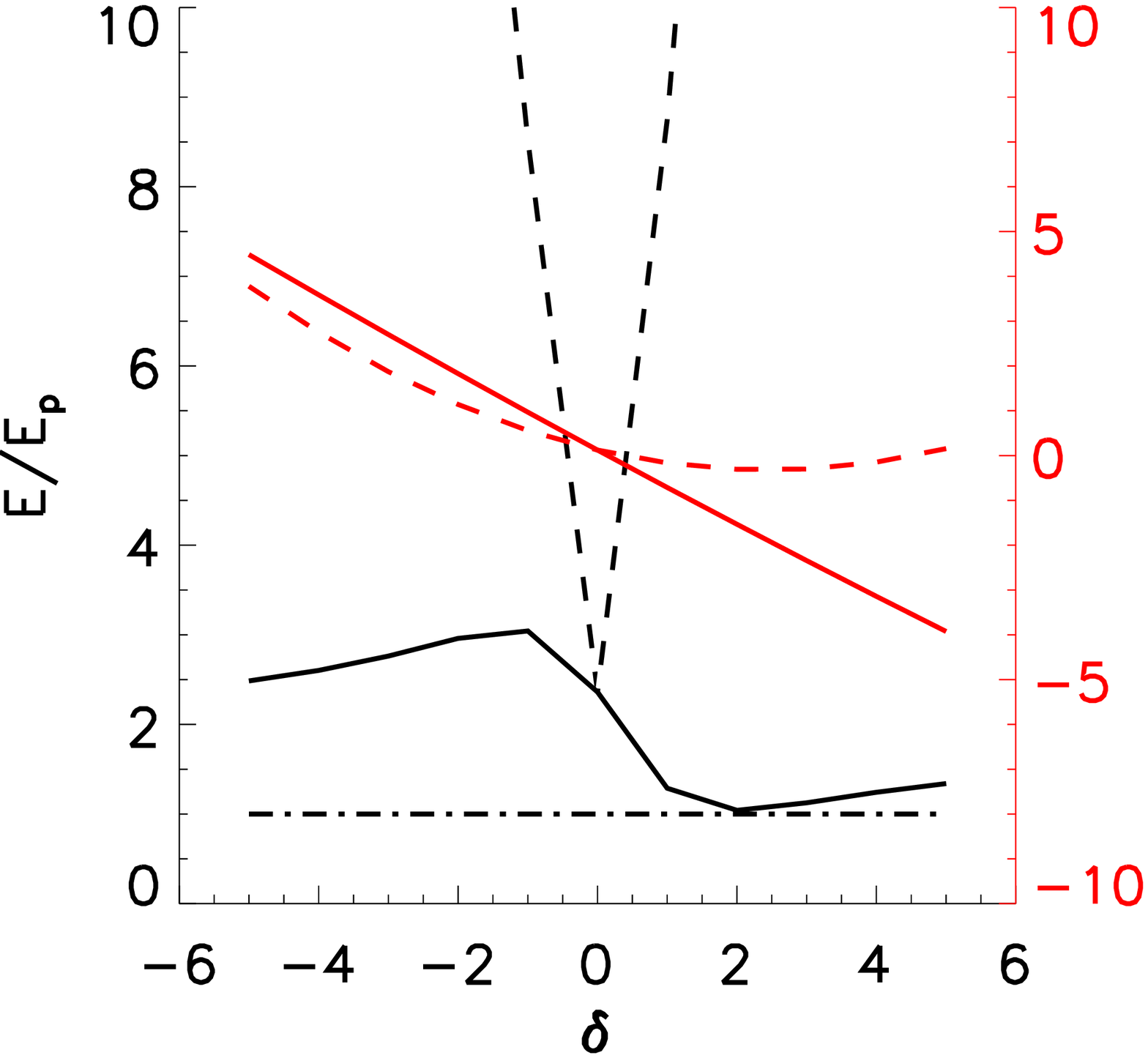}
   \includegraphics[width=0.33\textwidth]{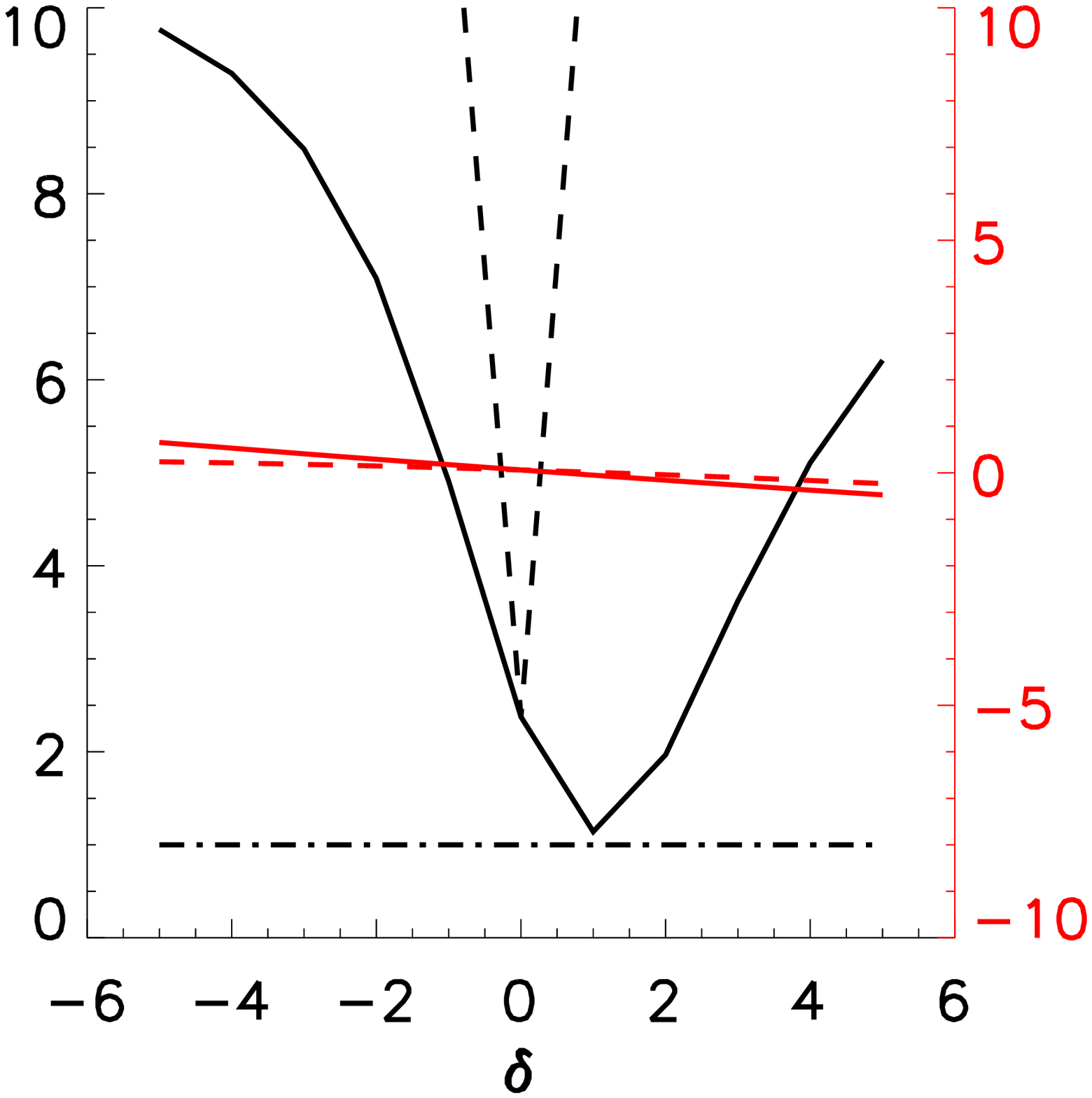}
   \includegraphics[width=0.33\textwidth]{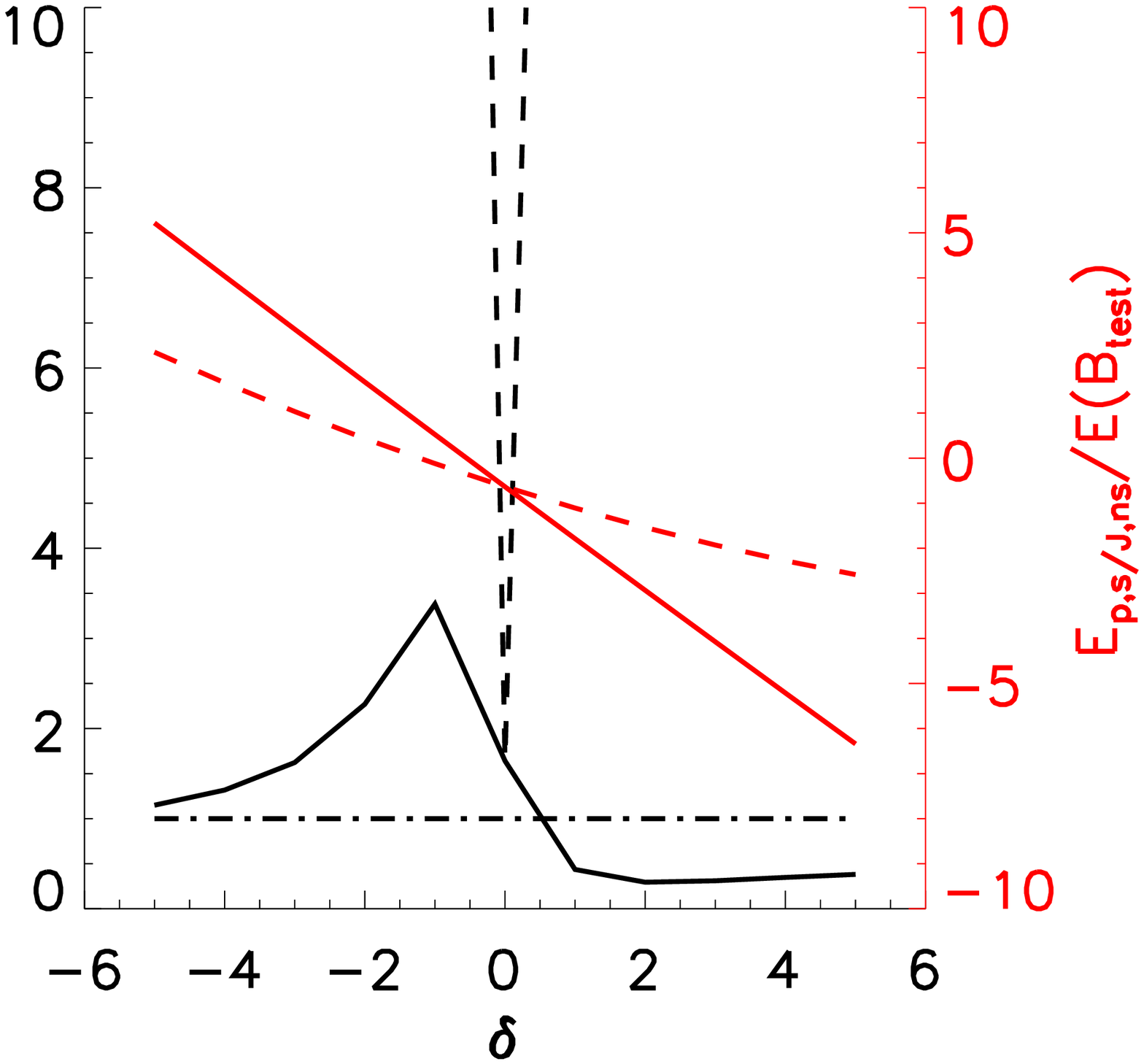}
                  }
 \raisebox{0pt}[0pt][0pt]{\raisebox{87mm}{\textcolor{black}{\textbf{
       \hspace{6mm} a) $\BDD$ \hspace{35mm} b) $\BTD$  \hspace{35mm} c) $\BMHD$  }}}}
 \raisebox{0pt}[0pt][0pt]{\raisebox{50mm}{\textcolor{black}{\textbf{
        \hspace{6mm} d) $\BEXsq$ \hspace{35mm} e) $\BEXfep$ \hspace{32mm} f) $\BEXfe$  }}}}
 \vspace{-8mm}
 \caption{Magnetic energy, \eq{enns}, normalized to the energy of the corresponding potential field  (black lines, zoom of Figure~\ref{f:endivb_sr}), and  energy associated to the nonsolenoidal part of $\vB_J$, normalized to the energy of $\vBt$  (red lines, equal to $\EdivBJ / E$ in the notation of \eq{emixprime}), as a function of the amplitude $\delta$ of the nonsolenoidal term, \eq{btest}.
Two models for the divergence are shown according to \eq{divmodels}.  Each panel shows the results of a test field:
(a) the potential field of a double dipole, $\BDD$;
(b) the TD model, $\BTD$;
(c) the MHD model, $\BMHD$;
(d) the NLFFF model of the nonpreprocessed magnetogram of AR~11158, $\BEXsq$;
(e) the NLFFF model of the preprocessed magnetogram of AR~11024, $\BEXfep$;
(f) the NLFFF model of the nonpreprocessed magnetogram of AR~11024, $\BEXfe$.
The black dash-dotted line at $E/\Ep=1$ marks the value below which the solution is unphysical.
A large change of scale of both axes is present between the top and the bottom rows.
}
 \label{f:endivb_ur}
\end{figure*}

On the other hand, Model~1 shows a more complex dependence on $\delta$, which is shown in magnified scale by the black lines in Figure~\ref{f:endivb_ur}.
Counterintuitively, the largest variation in the relative energy $E / \Ep$ as a function of $\delta$ is found for  $\BEXfep$, \ie for the extrapolation case, which satisfies Thomson's theorem better, see  Table~\ref{t:thompson}.
The continuous black lines in  Figure~\ref{f:endivb_ur}d,f show the presence of one maximum and one minimum in the considered range of values of $\delta$ (for the $\BEXfep$ case, these lie outside the considered range), implying that, at high values, the potential field energy grows faster than the total energy.
The location of the extrema is different in the three $\BEXsq$,  $\BEXfep$, and  $\BEXfe$ cases, and in none of the cases are the extrema found for the solenoidal ($\delta=0$) or the test ($\delta=1$) configurations. 
In general, the maximum and minimum energy configurations depend on the spatial distribution of the divergence of the test field, through $\Esns$. 
\subsection{Unphysical cases.}\label{s:unphysical}
The black dash-dotted line at $E/ \Ep=1$  in Figure~\ref{f:endivb_ur}  is the value below which unphysical fields are obtained.
We find that only Model~1 can produce unphysical solutions, and only for specific range of $\delta$ values in the  $\BDD$, $\BTD$,  $\BEXfe$ cases. 
The latter case is known from the value of $\Ep/E$ in Table~\ref{t:thompson}, and is considered to be an extreme case because of the large divergence that it involves.
However, the possibility of also creating unphysical solutions in the far more solenoidal field $\BDD$ and $\BTD$ (for values of $|\delta|>5$) is unexpected.
It confirms that not just the value of the divergence is important, but also its detailed spatial distribution with respect to the solenoidal component, as evident from  $\Esns$. 
It is the alignment between $\vBdiv$ and $\vBs$, and not just the magnitude of $\vBdiv$,  which determines how strongly the energy depends on $\delta$. 
Moreover, while $E > \Ep$ is always satisfied for  $\BEXsq$ and  $\BEXfep$, the minimum value of $E$ is close to $\Ep$ (see  Figure~\ref{f:endivb_ur}d,e), showing that unphysical fields may be found relatively easily in NLFFF extrapolations.

From Table~\ref{t:emix_gi} and the related discussion of \sect{contrib_Emix} we showed that the main source of violation of Thomson's theorem is the term $E_{\rm p,s/J,ns}$ in $\Emix$.
The dependence on $\delta$ of this term, normalized to the energy of the test field, is shown by the red curves in  Figure~\ref{f:endivb_ur} for both models of divergence (\eq{divmodels}).
The contribution to the total energy is negligible in the  $\BDD$ and $\BMHD$ cases, and can be a few percent for large $\delta$ in the $\BTD$ case.

In the extrapolated cases, the dependence of $E_{\rm p,s/J,ns}$ on $\delta$ is linear for Model~1 and parabolic for Model~2. 
In  Model~1, the steepness of the linear curve increases, going to  $\BEXfep$ to  $\BEXsq$ and  $\BEXfe$, as expected (see Table~\ref{t:thompson} and related text).  
The amplitude of the error is two to three orders of magnitude larger than in the $\BDD$, $\BTD$, and $\BMHD$ cases.
In the  $\BEXfep$ case the error is smaller, but it is still about a factor 20 larger than in $\BTD$ for $\delta=5$.

If we consider the black curves in  Figure~\ref{f:endivb_ur} for  $\delta=0$, we can identify the energy of the solenoidal field as a natural reference value for the free energy. 
Starting from this reference value, for increasing $|\delta|$, the linear contribution of  $E_{\rm p,s/J,ns}$, together with the quadratic change in the potential field energy, creates the maximum and minimum values of $E/ \Ep$.
If the linear contribution is large enough, the minimum lies below the threshold $E/ \Ep=1$, and there is a range of values where the solution is unphysical.
For even higher values of  $|\delta|$, the quadratic dependence of the potential field energy dominates $E$. 
From this point onward, $E_{\rm p,s/J,ns}$ is not the main source of error in \eq{kelvin}.

More generally, a parametric study like the one in Figs.~\ref{f:endivb_sr} and \ref{f:endivb_ur} can be used to identify what is the level of divergence (\ie the level of $\Emix$or  $E_{\rm p,s/J,ns}$) that can be tolerated and which is the threshold above which the solution becomes entirely unphysical (\ie with $E/ \Ep<1$).

In conclusion, the parametric study shows that the energy may be severely influenced by the solenoidal property of the field.
The effect depends not only on the amplitude of the nonsolenoidal component, but also on 
the specific average orientation of the nonsolenoidal component with respect to the solenoidal one (directly affecting $\Esns$ in \eq{enns}).
As a result, a single-number divergence metric, such as $\avfi{}$, is insufficient to deduce what errors should be expected in the energy. 
A more proper indication is found by the numerical verification of Thomson's theorem (\sect{test_Thompson}) and by a parametric study as presented in this section.

\section{Source of divergence in NLFFF extrapolations}
\label{s:divestrap}
We now investigate in more detail some of the test fields discussed in \sect{test_Thompson}, with emphasis on the reason for the  large divergence that leads to violating  Thomson's theorem.
The main source of error comes, in almost all the cases, from the mixed term $\Emix$, and is associated with the nonsolenoidal component of the current-carrying part of the field.
Also, there are markedly larger errors in the extrapolated test fields, $\BEXsq$, $\BEXfep$, $\BEXfe$, than in $\BDD$, $\BTD$, and $\BMHD$.
Finally, the preprocessing of the vector magnetogram before extrapolation yields more solenoidal fields, whereas a simple averaging does not seem to be enough for removing errors, and yields a more severe violation of Thomson's theorem (\eq{kelvin_exact}).
\subsection{Analysis of small scales}\label{s:smallscales}
\begin{figure*}
\centering
\sidecaption    
\includegraphics[bb=0 50 516 520, width=6cm]{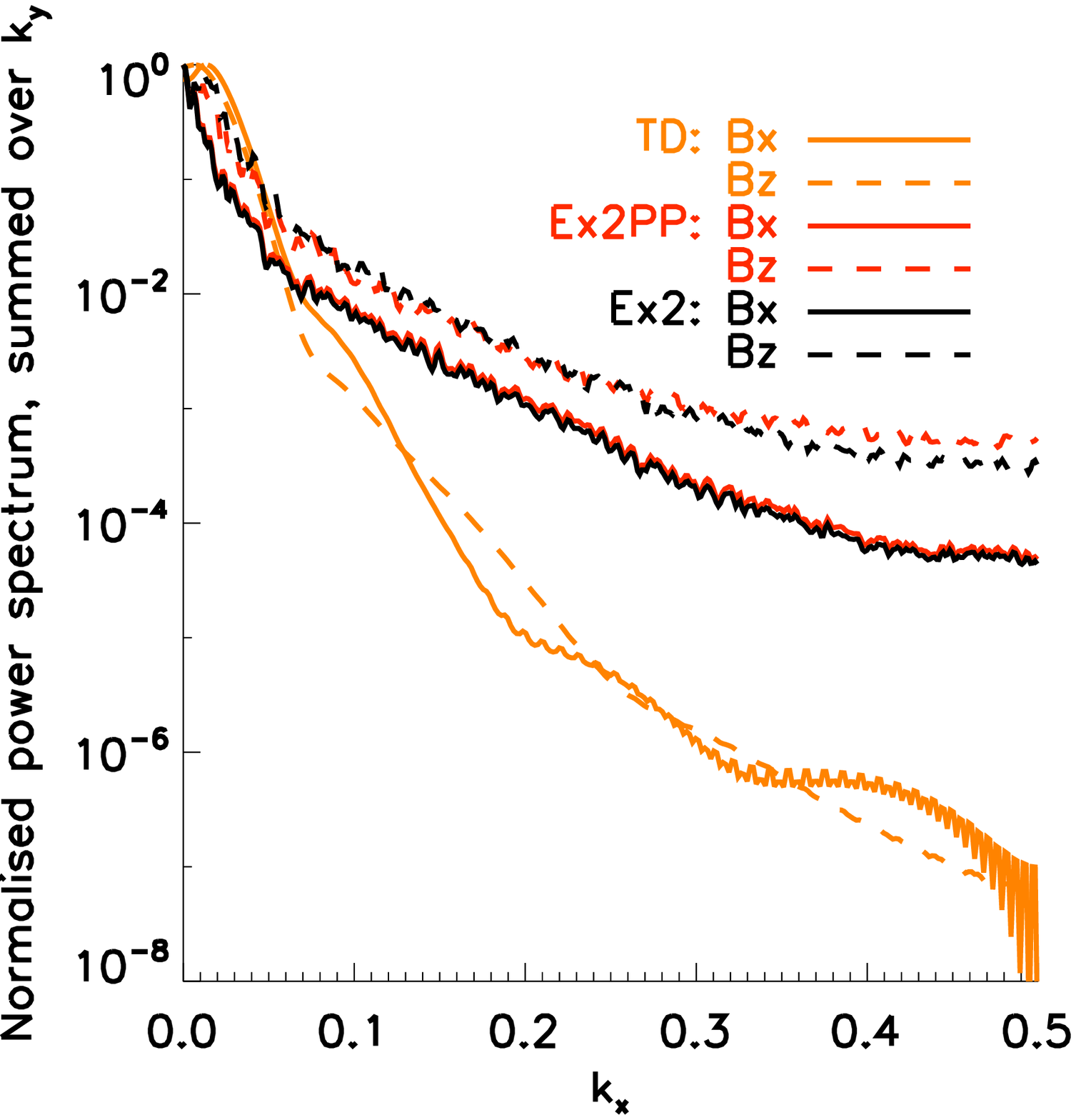}
\includegraphics[bb=0 50 516 520, width=6cm]{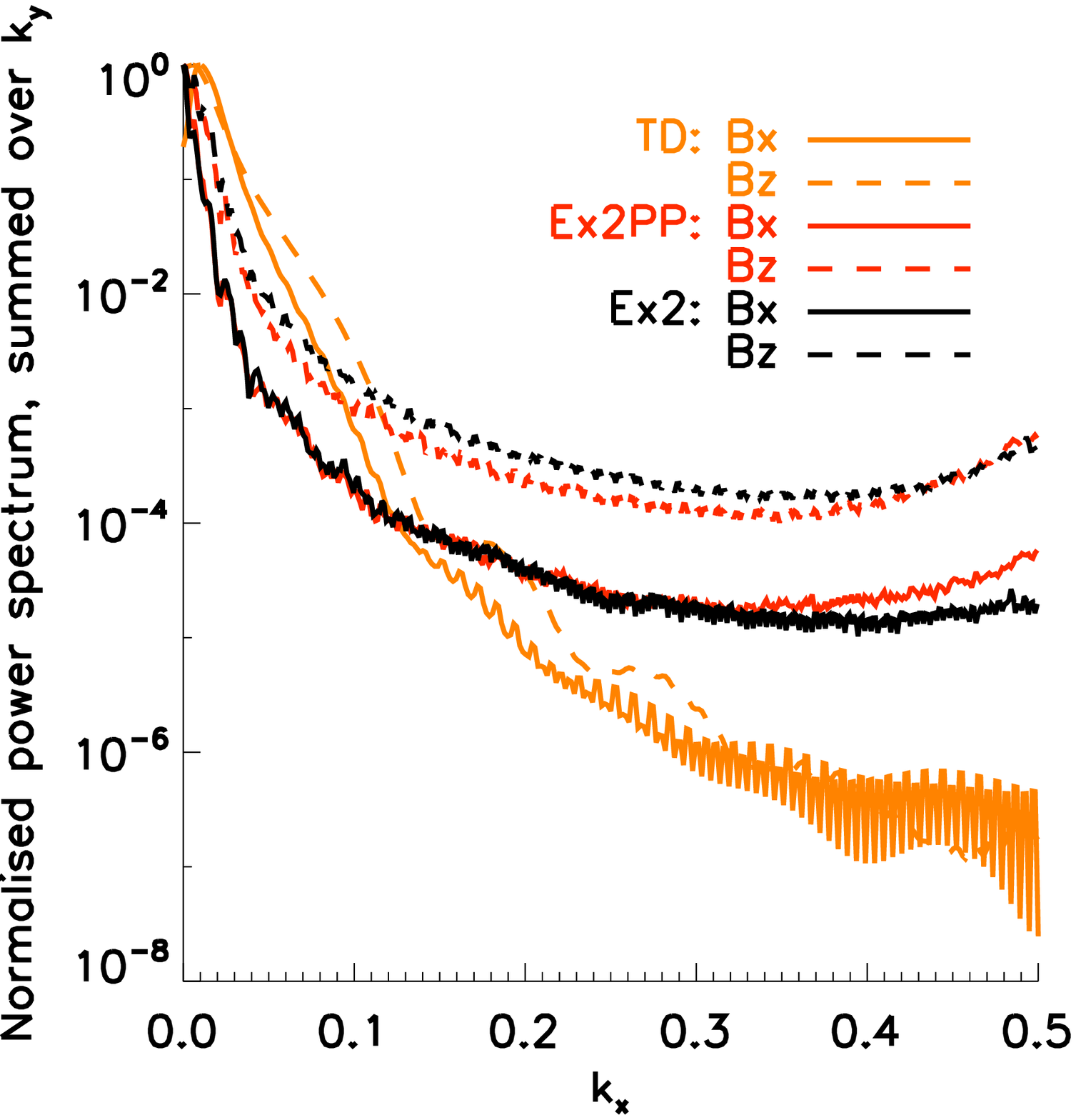}
 \caption{Power spectra of the two-dimensional fields $B_x$ (continuous line) and $B_z$ (dashed line), summed over all wave numbers $k_y$, for the three cases $\BTD$ (orange), $\BEXfep$ (red), and $\BEXfe$ (black).
\textbf{Left}: at the bottom boundary. \textbf{Right}: at the tenth pixel in height. 
Spectra are normalized to their maximum value, and the spatial resolution is taken to be 1 in both directions (\ie the wave number $k_x$ has the dimension of pixel$^{-1}$ and is normalized to the total number of modes).}
\label{f:spectra}
\end{figure*}

One main difference among the $\vBt$ cases in the upper half of Table~\ref{t:thompson} is the length scale of the magnetic field: While the first three cases are smooth fields with a magnetic field variation spanning several times the spatial resolution, the extrapolated cases have large variations on the pixel scale, especially  at the bottom boundary, \ie on the vector magnetogram that is used as a boundary condition for extrapolations. 
This is true to a different degree for the three cases: For $\BEXsq$ the vector magnetogram was interpolated (with a flux-conserving average) at a resolution of about one third that of $\BEXfe$ and $\BEXfep$. 
Such an interpolation smooths part of the small scale away, yielding results that are closer to the $\BEXfep$ case rather than to the $\BEXfe$ one.
$\BEXfep$ is not interpolated, but it is preprocessed, an operation that includes an explicit smoothing of smaller scales, especially on the transverse components.
Finally, $\BEXfe$ has neither interpolation nor preprocessing, and it retains all the small scales that are present at the full resolution of the \textit{Hinode}/SOT vector magnetograms.

As an example, \fig{spectra} shows the power spectrum of the $x$- and $z$-components of the fields  $\BTD$,  $\BEXfep$, and  $\BEXfe$, at two different heights as a function of the normalized wave number $k_x$.
The lefthand panel of the figure shows that, at the bottom boundary, $\BTD$ has power spectra that decrease rapidly with $k_x$, in both components. 
In contrast, the power spectra of $\BEXfep$ and $\BEXfe$ have higher values on all scales, which are particularly strong in the vertical component.

Ten pixels above the bottom boundary (right panel in \fig{spectra}), the $\BTD$ power spectrum is essentially the same as at $z=0$ because both planes cut through the flux rope, so a similar magnetic structure is present.
In contrast, $\BEXfep$ on the upper plane has a much more peaked spectrum, except for the distribution tail on the smallest scales which is basically as strong as at the bottom boundary.
Such a component on the shortest scales comes from the force-free condition that is enforced by the extrapolation code, which  propagates into the volume the small scales that are present at the bottom boundary. 

We now consider the difference between preprocessed case $\BEXfep$ and the non-preprocessed one $\BEXfe$.
The difference in $\avfi{}$ between the two is about a factor 2, and it is large in the other energy metrics in Table~\ref{t:thompson}. 
The comparison between the normalized spectra of $\BEXfep$ and $\BEXfe$ in  \fig{spectra} shows that there are comparable (relative) energies on small scales in both cases. 
 Actually, by locally changing the magnetic field at the bottom boundary to enforce force-free compatibility, preprocessing increases the small scales.
The smoothing term that is present in preprocessing only has a limiting effect on such an increase.
Therefore, the two cases  $\BEXfep$ and $\BEXfe$ do not differ strongly as far as the presence of small scales is concerned, while Thomson's theorem is much better satisfied for $\BEXfep$ than for $\BEXfe$ (see \sect{ThompsonTest}).

The cleaned test fields $\vBts$ are numerically solenoidal, and there is no violation of Thomson's theorem.
However, in these cases, too, small scales are increased (not shown), since the divergence cleaner introduces extra electric currents that are related to derivatives of the divergence of the original field, see \eq{jcleaner}. 
This is an additional confirmation that the presence of small scales as such is not \textit{directly} at the origin of the violation of Thomson's theorem.

\subsection{Role of small scales and preprocessing}\label{s:rolescales}
\citet{2010A&A...519A..44V} show that the NLFFF  extrapolation of the $\BTD$ vector magnetogram yields a very accurate reconstruction of the whole test field, which is also solenoidal to a very high degree.
On the other hand,  there is a large difference in the scale distribution between smooth fields like the $\BTD$ and the extrapolated fields.

The presence of small scales inside the volume, which are induced by the small scales at the boundary, may not be correctly approximated by the discretization employed in extrapolation code, yielding local violation of the solenoidal constrain.
However, when the extrapolation from a preprocessed magnetogram is considered, the extent of the violation of Thomson's theorem is greatly reduced, even though small scales are actually increased. 
By partially enforcing force-free compatibility on the bottom boundary, the preprocessing provides the extrapolation code with a boundary condition that is more compatible with the force-free equations. 
Since extrapolation codes attempt to construct a solution of the force-free equations that is simultaneously force- and divergence-free, the more compatible the boundary, the more consistent (\ie force- and divergence-free) the obtained solution.
Conversely, when the boundary condition is incompatible with the force-free equation,  the reduction of the Lorentz forces is at the expense of the solenoidal condition. 
In such cases, the divergence of the solution is higher, and Thomson's theorem is more severely violated.
We thus conclude that the incompatibility of the boundary condition with the force-free condition is at the origin of the difference in the errors $\EdivBJ$ and $\Emix$ between $\BEXfep$ and $\BEXfe$.

We notice that preprocessing is a parametric method that can produce progressively more force-free-compatible vector magnetograms for higher values of the employed parameters, at the price of larger modifications of observed values. 
The energy values and their relative errors therefore vary continuously as a function of the preprocessing parameters, quite independently of the particular extrapolation method that is employed \citep[see, \eg][]{2008ApJ...675.1637S,2008SoPh..247..269M}. 
No unequivocal method is available in order for determining the best parameters to use \citep[see, \eg][]{2006SoPh..233..215W,2011A&A...526A..70F,2012SoPh..281...37W}, which leaves energy estimations subjected to uncomfortable arbitrariness.

\section{Conclusions}\label{s:conclusions}

Thomson's theorem states that the energy of a magnetic field is given by the sum of the energy of the current-carrying part of the field plus the energy of the potential field that has the same distribution of the normal component on the boundary of the considered volume. 
The field must be perfectly solenoidal for the theorem to be valid. 
Such a condition is often only approximately satisfied in numerical simulations, such as in MHD simulations and NLFFF extrapolations.
However, it is a non-trivial task to identify a quantitative estimation of solenoidal errors that can be applied to different discretizations of magnetic fields, essentially due to the non-local consequences that such errors produce.
Our goal has been to develop physically meaningful metrics and practical methods that can be used to judge whether the solenoidal property is fulfilled with sufficient accuracy. 

To this aim, we introduced a decomposition of the energy of a discretized field into solenoidal and nonsolenoidal contributions that allowed an unambiguous and numerically well-defined estimation of the effect of the divergence in terms of associated energies.
Moreover, we introduced a method of parametrizing the divergence that allows for an exploration of the nonsolenoidal effects.

In this way, the numerical verification of Thomson's theorem offers an operational and quantitative way of checking the reliability of energy estimations in numerical computations.
Since the violation of Thomson's theorem is solely determined by the presence of magnetic charges, it is at the same time a quantitative estimation of the importance of solenoidal errors.

We applied our method to six different test cases, covering a representative sample of numerical realizations.
Of the six test cases considered here, two of them (the dipolar field $\BDD$ and a snapshot of an MHD simulation of null-point reconnection $\BMHD$) presented negligible violations, and one (a force-free current ring $\BTD$) offered only a moderate one that, however, has finite effects on the energy. 
In the case of an NLFFF extrapolation of a preprocessed vector magnetogram ($\BEXfep$), the sum of the potential energy $\Ep$ and free energy $\EJs$ is very close to the total energy $E$, and one could draw the conclusion that almost no violation of Thomson's theorem occurs.
However, by separating all contributions in \eq{kelvin}, our analysis showed compensating energy contributions ($\EdivBJ$ and $\Emix$) that are close to $\EJs$.
If the most conservative view is adopted by considering errors in absolute values, then the opposite conclusion must be drawn: The violation is large enough to compromise the estimation of the free energy, since both $\EdivBJ$ and $\Emix$ are on the order of the free energy value $\EJs$. 
The last two cases we studied, also NLFFF extrapolations but of nonpreprocessed magnetograms ($\BEXsq$ and  $\BEXfe$), represent cases with very large errors.

The energy of the potential field $\Eps$ is the reference value for the free energy.
In our applications, the inaccuracy in its determination, $\EdivBp$,  is practically never significant.
The current-carrying part of the field is responsible for the largest errors instead.

The parametric study shows that the amplitude of the nonsolenoidal component is not the only factor that generates errors in the energy. 
The average orientation of the nonsolenoidal component with respect to the solenoidal one (affecting directly $\Esns$ in \eq{enns}) plays an even more important role.
Indeed, even using a relatively solenoidal discretized magnetic field (like $\BTD$), it is possible to create configurations where the energy of the field is lower than that of the corresponding potential field. 
Such unphysical solutions have also been found in some cases of NLFFF extrapolations. 

More generally, in NLFFF extrapolations the energy of the reconstructed field was found to vary according to the extent of the modification that was enforced on the vector magnetogram that is used as boundary condition (by preprocessing, \ie by smoothing and/or by enforcing force-freeness compatibility). 
Our study shows quantitatively the effect of such practices on the energy, and makes it clear that the origin of the variability (and errors) in energy estimations based on NLFFF extrapolations is the presence of a large divergence, which is eventually caused by the lack of compatibility between the equations solved (solenoidal force-free field) and the photospheric boundary conditions, rather than by noise or the small scales present in the vector magnetogram.

Finally, the parametric study is based on a numerically solenoidal field that is obtained from a given, nonsolenoidal one. 
We introduced a method for the complete removal of the nonsolenoidal component of a discretized field.
At the price of changing boundary values and the current density, this method provides a field that is solenoidal to numerical precision. 
When the solenoidal versions of the test fields are considered, the Thomson theorem is found to be fulfilled with more than 99\% accuracy.

We concluded that testing Thomson's theorem in numerical realizations of magnetic fields is a powerful method quantifying the amount of nonsolenoidal contributions to a numerical magnetic field.
In particular, it allows assessing the reliability of free magnetic energy estimations, a crucial quantity in phenomena such as flares and coronal mass ejections.
To this purpose, we proposed a set of analytical and numerical tools that allowed us to fully test the reliability of numerical magnetic fields.
Such a set includes a method for removing the divergence from a given discretized field, to numerical precision. 
The effect of larger and larger divergence contributions is studied  by parametrically adding a known divergence to the numerically solenoidal field. 
In this way, it is possible to monitor the effect of the nonsolenoidal part of the magnetic field and to quantify its effect in terms of  magnetic energy.
Our method can be applied to any discretization of magnetic fields, \eg in MHD simulations and in NLFFF extrapolations, to constrain quantitatively errors due to violation of the solenoidal property.
 
\begin{acknowledgements}
The authors are pleased to thank Guillaume Aulanier for fruitful discussions, Bernhard Kliem and Tibor T\"or\"ok for making the numerical solution of the TD equilibrium available, and the referee for helpful comments that improved the clarity of the paper.
GV is indebted to the NLFFF Consortium for stimulating discussions and collaborations. 
The research leading to these results has received funding from the European Commission's Seventh Framework Program (FP7/2007-2013) under the grant agreement eHeroes (project n$^\circ$~284461, www.eheroes.eu).
SM gratefully acknowledges support from the NASA Postdoctoral Program, administrated by Oak Ridge Associated University through a contract with NASA, during her stay at NASA Goddard Space Flight Center.
Calculations were done on the quadri-core bi-Xeon computers of the Cluster of the Division Informatique de l'Observatoire de Paris (DIO). 
SDO data are courtesy of NASA/SDO and the HMI science team.
Hinode is a Japanese mission developed and launched by ISAS/JAXA, collaborating
with NAOJ as a domestic partner, NASA and STFC (UK) as international partners. Scientific operation of the Hinode mission is conducted by the Hinode science team organized at ISAS/JAXA. This team mainly consists of scientists from institutes in the partner countries. Support for the post-launch operation is provided by JAXA and NAOJ (Japan), STFC (UK), NASA (USA), ESA, and NSC (Norway). 
\end{acknowledgements}

 \bibliographystyle{aa}

\begin{appendix} 
\section{Details of the numerical implementation}\label{s:num_details}
In the applications presented in this paper we have considered uniform Cartesian grids of resolution $\Delta$ in all directions, discretizing a rectangular volume $\vol$ (see \sect{testfields} for the actual values of $\Delta$ and $\vol$ in each case).
We compute derivatives using  the standard second-order, central-difference operator, and we employ the relevant one-sided (\ie forward or backward), second-order differences at the boundaries of $\vol$. 
The only exception is the computation of the divergence of $\vBt$, since all test fields are known in a volume that is larger than the selected $\vol$ (on lateral and top boundaries). 
In this case,  $\divBt$ is computed using the central differences also at the location of the lateral and top boundaries of $\vol$.

In the computation of volume integrals,  the cell volume $\Delta^3$ is assigned to each internal node of the grid, whereas the cell volume is reduce to half, one fourth, and one eighth for nodes on the lateral surfaces, edges, and corners of $\vol$, respectively. 
Similarly, in the computation of surface integrals, the cell surface  $\Delta^2$ is assigned to each node inside each side of $\vol$, whereas the cell surface is reduced to half and one fourth on edges and corners of each side, respectively.
Despite the accurate computation of integrals, the divergence theorem, \eq{divTheorem}, is not insured to hold numerically, a property that requires special techniques, like finite-volume discretizations, to be fulfilled. 
\section{Divergence cleaner} \label{s:cleaner}
To construct a numerically solenoidal field [$\vBs$] from a field [$\vB$] let us define 
  \BE 
  \vBs=\Nabla \times \vA \p{\,,}
  \label{eq:brota}
  \EE
where $\vA$ is the vector potential computed from $\vB$ in the volume $\vol=[x_1,x_2]\times[y_1,y_2]\times [z_1,z_2]$. 
The vector potential $\vA$ can be derived as in \citet{2012SoPh..278..347V} using the gauge $\hatz\cdot\vA=0$, yielding the expression
  \BE 
  \vA = \vb+\hatz\Times \int_{z}^{z_2}  \vB \ \rmd z' \, ,
  \label{eq:afina}
  \EE
where $\vb \equiv (\Ax (x,y,z=z_2),\Ay (x,y,z=z_2),0)$ is any solution of 
  \BE 
  0 = \partial_x \by -\partial_y \bx -B_{\rm ns,z} (x,y,z=z_2) \, . 
  \label{eq:acoeff}
  \EE
A direct substitution of \eq{afina} into \eq{brota} shows that
  \BE 
  {\vB}_{\rm s} \equiv \curlA = \vB+\hatz\int_{z}^{z_2} (\divB) \ \rmd z'\,, 
  \label{eq:rota}
  \EE
with the property that $\Nabla \cdot {\vBs}=0$.  
In other words, \eq{rota} naturally separates $\vB$ into a solenoidal part ${\vB_{s}}$ and a nonsolenoidal one, thus defining a divergence cleaner for $\vB$.
The $z$-component of $\vB$ is changed throughout the volume except on the top boundary, whereas the $x-$ and $y$-components are unchanged. 
The amplitude of the modification to $\vB$ at a given height $z$ is given by the cumulative effect of ``magnetic charges'' above that altitude.

Since only the $z$-component of the field is changed, the divergence cleaner changes the $x$- and $y$-components of the current, but not the $z$-component,
  \BE 
  \vec{J}_{\rm s}= \vec{J}+(\partial_y,-\partial_x,0)\int_{z}^{z_2}  (\divB) \ \rmd z'\, ,
  \label{eq:jcleaner}
  \EE
therefore the cleaner changes the injected magnetic flux but not the injected electric current through the bottom layer. 
On the other hand, since most of the test fields considered in this article have the highest values of divergence close to the bottom boundary, only the lower part of the field is changed significantly by the cleaner.

Computation $\vBs$ requires numerical computation of an integral of the type $G(z)=\int_{z}^{z_2} \ f(t) \, \rmd t$, as in \eq{rota} for $f=\Nabla \cdot \vB$. 
To achieve numerical accuracy in the solenoidal property of $\vBs$, $G(z)$ must satisfy $\partial_z G(z)=-f(z)$ numerically, \ie must satisfy the numerical formulation of the fundamental theorem of integral calculus in the employed discretization.
For the second-order central differences that are used in the analysis, this can be obtained by the recurrence formulae 
  \BA 
  G(n_{\rm z}-1) &\equiv& 0    \, , \nonumber \\
  G(k) &=& G(k+2)+2 \Delta \, f(k+1)\,,   \quad 0 \le k \le  n_{\rm z}-3,    \label{eq:recurrence}
  \EA
where $G(z)=G(z_1+k\Delta)\equiv G(k)$ with $k=0,1,2, \cdots, (n_{\rm z}-1)$,  and $\Delta$ is the uniform spatial resolution in $z$. 

The constraint $\partial_z G(z)=-f(z)$ in the second-order, central-difference discretization does not fix the value of $G(n_{\rm z}-2)$.
To do that, we require that the divergence of \eq{rota} also vanishes at the bottom boundary, \ie $(\Nabla \cdot \vBs)|_{\rm{z=z_1}}=0$.
Here the second-order divergence operator is computed by using a second-order, forward derivative in the $z$-direction, \ie defining the operator $ \Nabla_{\rm os}\equiv \Nabla_{\rm x,y} + \hatz\partial_z^{\rm os}$, where $\Nabla_{\rm x,y}\equiv \hatx\partial_{\rm x}+\haty\partial_{\rm y}$ and $(\partial_z^{\rm os}f)(0)=(-3f(0)+4f(1)-f(2))/2\Delta$.
By using the recurrence formula \eq{recurrence}, the condition on the bottom boundary is transformed into the condition for $G(n_{\rm z}-2)$, yielding 
\BE
  G(n_{\rm z}-2) = \Delta \, \left[ \, 2 
      \left(\sum_{\rm{even \ k}=2}^{n_{\rm z}-2} - \sum_{\rm{odd \ k}=1}^{n_{\rm z}-3} \right)
      f(k) +\, \frac{1}{2}\, [ f_{\rm os}(0) 
                                     - 3f(1) ] \,\right] \,, \nonumber 
\EE
where $f_{\rm os}=\Nabla_{\rm os} \cdot \vB$.
Such a numerical trick is only possible if the volume is discretized by an even number of points in the $z$-direction, therefore the analysis volumes employed in the article were chosen to satisfy such a requirement.
\section{Measures of $\divB$}\label{s:numdivb}

The total divergence of a field $\vB$ can be conveniently expressed by a single number using the average $\avfi{}$ over the grid nodes of the fractional flux
  \BE
  f_i \equiv \frac{\int_{v} d{v} \ (\Nabla \cdot \vB)_i}{\int_{\partial{v}} dS |\vB_i|},
  \label{eq:fi}
  \EE
through the surface $\partial{v}$ of a small volume ${v}$ including the node $i$ \citep{2000ApJ...540.1150W}.
Taking a cubic voxel of side equal to $\Delta$ as the small volume ${v}$ centered on each node, the divergence in the discretized volume $\vol$ of uniform and homogeneous resolution $\Delta$ is then given by
  \BE
  \avfi{}=\frac{\Delta}{6N}\sum_{i}\frac{|\divB_i|}{|\vB_i|} \p{\,,}  
  \label{eq:<fi>}
  \EE 
where $i$ runs over all $N$ nodes in $\vol$.
This metric depends on the considered volume, so that values are strictly comparable only if computed on equal volumes. 

\end{appendix}

\end{document}